\documentclass[aps,pra,longbibliography,notitlepage,reprint,twocolumn,amsmath,amssymb,superscriptaddress]{revtex4-1}

\usepackage{graphicx}
\usepackage{bm}
\usepackage{braket}
\usepackage{natbib}
\setcitestyle{numbers,square,sort&compress}

\usepackage{here}

\usepackage{times}
\usepackage[english]{babel}

\graphicspath{{pict/}{}}

\newcommand\hrefBibPDF[3][]{}

\usepackage{hyperref}
\hypersetup{pdfstartview={FitH},pdfpagemode={UseNone},
	colorlinks,linkcolor=blue, citecolor=blue, urlcolor=blue,
	bookmarksopen=true, pdfnewwindow=true}

\usepackage[all]{hypcap}

\usepackage[nodocdata=true]{pdfprivacy}

\begin{document}

\title{Quantum metasurface for multi-photon interference and state reconstruction}

\author{Kai~Wang}
\affiliation{Nonlinear Physics Centre, Research School of Physics and
	Engineering, The Australian National University, Canberra, ACT 2601, Australia}

\author{James~G.~Titchener}
\affiliation{Nonlinear Physics Centre, Research School of Physics and
	Engineering, The Australian National University, Canberra, ACT 2601, Australia}
\affiliation{Quantum Technology Enterprise Centre, Quantum Engineering Technology Labs, H. H. Wills Physics Laboratory and Department of Electrical and Electronic Engineering, University of Bristol, BS8 1FD, UK}

\author{Sergey~S.~Kruk}
\affiliation{Nonlinear Physics Centre, Research School of Physics and
	Engineering, The Australian National University, Canberra, ACT 2601, Australia}

\author{Lei~Xu}
\affiliation{Nonlinear Physics Centre, Research School of Physics and
	Engineering, The Australian National University, Canberra, ACT 2601, Australia}
\affiliation{School of Engineering and Information Technology, University of New South Wales, Canberra, ACT 2600, Australia}

\author{Hung-Pin~Chung}
\affiliation{Nonlinear Physics Centre, Research School of Physics and
	Engineering, The Australian National University, Canberra, ACT 2601, Australia}
\affiliation{Department of Optics and Photonics, National Central University, Jhongli 320, Taiwan}

\author{Matthew~Parry}
\affiliation{Nonlinear Physics Centre, Research School of Physics and
	Engineering, The Australian National University, Canberra, ACT 2601, Australia}

\author{Ivan~I.~Kravchenko}
\affiliation{Center for Nanophase Materials Sciences, Oak Ridge National Laboratory, Oak Ridge, Tennessee 37831, USA}

\author{Yen-Hung~Chen}
\affiliation{Department of Optics and Photonics, National Central University, Jhongli 320, Taiwan}
\affiliation{Center for Astronautical Physics and Engineering, National Central University, Jhongli 320, Taiwan}

\author{Alexander~S.~Solntsev}
\affiliation{Nonlinear Physics Centre, Research School of Physics and
	Engineering, The Australian National University, Canberra, ACT 2601, Australia}
\affiliation{School of Mathematical and Physical Sciences, University of Technology Sydney, Ultimo, NSW 2007, Australia}

\author{Yuri~S.~Kivshar}
\affiliation{Nonlinear Physics Centre, Research School of Physics and
	Engineering, The Australian National University, Canberra, ACT 2601, Australia}

\author{Dragomir~N.~Neshev}
\affiliation{Nonlinear Physics Centre, Research School of Physics and
	Engineering, The Australian National University, Canberra, ACT 2601, Australia}

\author{Andrey~A.~Sukhorukov}
\email[Corresponding author. Email: ]{Andrey.Sukhorukov@anu.edu.au}
\affiliation{Nonlinear Physics Centre, Research School of Physics and
	Engineering, The Australian National University, Canberra, ACT 2601, Australia}

\date{\today}

\begin{abstract}
	
Metasurfaces based on resonant nanophotonic structures have enabled novel types of flat-optics devices often outperforming the capabilities of bulk components, yet these advances remain largely unexplored for quantum applications. We show that non-classical multi-photon interferences can be achieved at the subwavelength scale in all-dielectric metasurfaces. We simultaneously image multiple projections of quantum states with a single metasurface, enabling a robust reconstruction of amplitude, phase, coherence, and entanglement of multi-photon polarization-encoded states. One- and two-photon states are reconstructed through nonlocal photon correlation measurements with polarization-insensitive click-detectors positioned after the metasurface, and the scalability to higher photon numbers is established theoretically. Our work illustrates the feasibility of ultra-thin quantum metadevices for the manipulation and measurement of multi-photon quantum states with applications in free-space quantum imaging and communications.

\end{abstract}

\maketitle

The field of nanostructured metasurfaces offers the possibility of replacing traditionally bulky imaging systems with flat optics devices~\cite{Yu:2014-139:NMAT} achieving high transmission based on all-dielectric platforms
\cite{Kuznetsov:2016-846:SCI,Decker:2015-813:ADOM,Arbabi:2015-937:NNANO,Kruk:2016-30801:APLP,Genevet:2017-139:OPT,Mueller:2017-113901:PRL}.
The metasurfaces provide a freedom to tailor the light interference by coherently selecting and mixing different components on a sub-wavelength scale,
enabling polarization-spatial conversion~\cite{Bomzon:2001-1424:OL,Arbabi:2015-937:NNANO, Pors:2015-716:OPT, Mueller:2016-42:OPT, Maguid:2016-1202:SCI, Ding:2017-943:ACSP, Mueller:2017-113901:PRL} and spin-orbital transformation~\cite{Devlin:2017-896:SCI}.
Such capabilities motivated multiple applications 
for the regime of classical light, yet the metasurfaces 
have a potential to emerge as essential components for quantum photonics~\cite{Jha:2015-25501:PRL,Roger:2015-7031:NCOM,Lyons:1709.03428:ARXIV,Stav:1802.06374:ARXIV}.

The key 
manifestations of quantum light are associated with non-classical multi-photon interference, which is an enabling phenomenon for the transformation and measurement of quantum states. Conventionally, manipulation of multi-photon states is performed through a sequence of beam-splitting optical elements, each realizing quantum interference~\cite{Silverstone:2014-104:NPHOT,Bayraktar:2016-20105:PRA,Fakonas:2015-23002:NJP}.
Recent advances in nanotechnology enabled the integration of beam-splitters and couplers on tailored plasmonic structures~\cite{DiMartino:2014-34004:PRAP,Vest:2017-1373:SCI}, yet
material losses and complex photon-plasmon coupling interfaces
restrict the platform scalability.
We realize several multi-photon interferences in a single flat all-dielectric metasurface. 
The parallel quantum state transformations are encoded in multiple interleaved metagratings,
taking advantage of the transverse spatial coherence of the photon wavefunctions extending across the beam cross section. In the classical context, the interleaving approach was effectively used for polarization-sensitive beam splitting~\cite{Bomzon:2001-1424:OL,Pors:2015-716:OPT,Maguid:2016-1202:SCI,Ding:2017-943:ACSP}, yet it requires nontrivial development for the application to multi-photon states.

\begin{figure*}
	\begin{center}
		\includegraphics[width=0.7\textwidth]{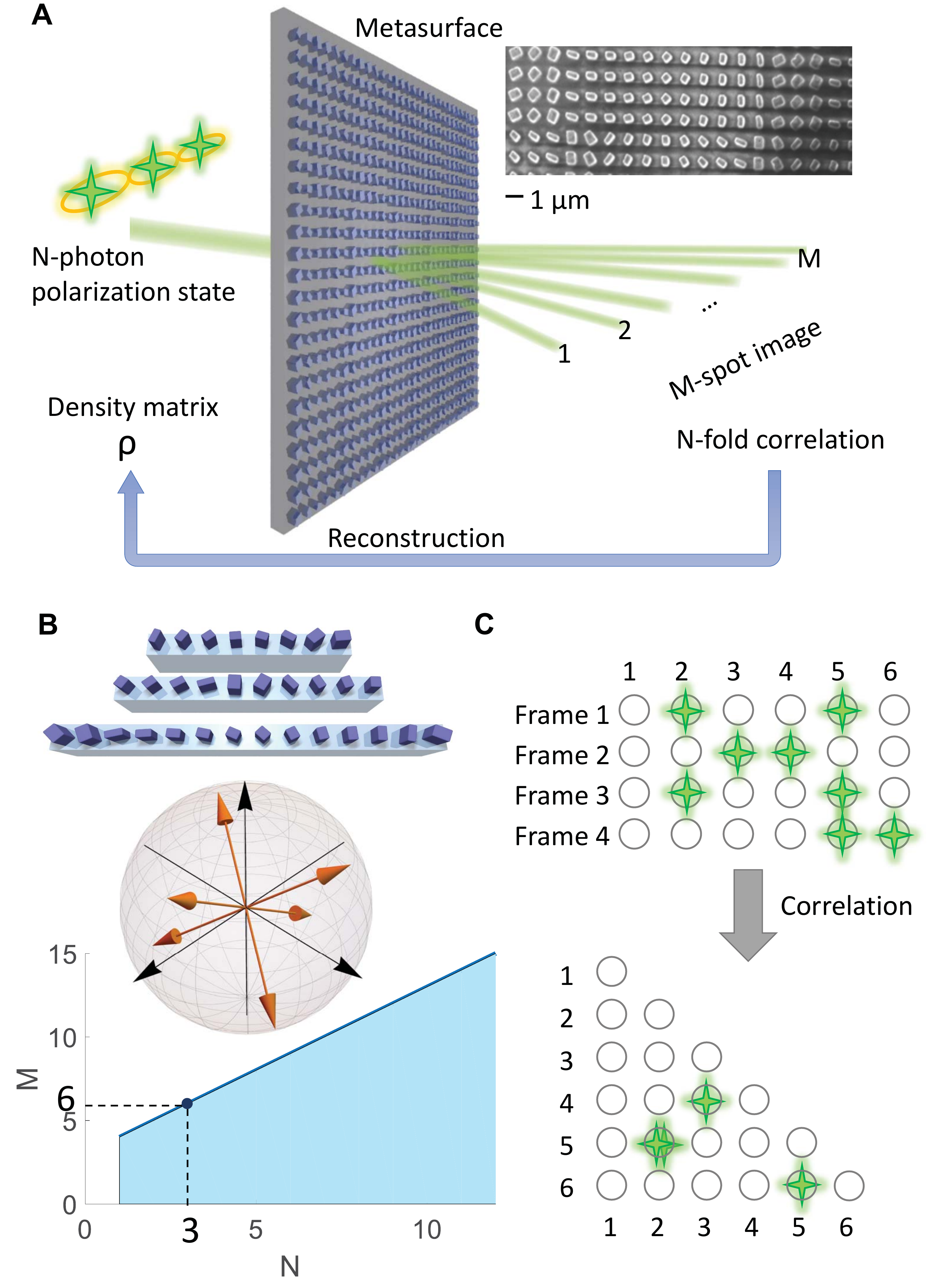}
		\caption{\textbf{Concept of quantum state imaging via nanostructured flat optics.} (A) Sketch of using a metasurface to image an input $N$-photon polarization state into an $M$-spot image. Top-right inset shows an SEM image of the fabricated all-dielectric metasurface. (B) Top~-- sketch of three interleaved gratings for $M=6$. Middle~-- the corresponding projective bases shown as vectors on the Poincar\'e sphere.  
Bottom~-- minimum number of required spots to fully reconstruct the initial quantum state for different $N$, where optimal-frame choice of projective bases exists for $M$=6, 8, 12, 20, $\ldots$.
(C) An example of correlation measurement with $N$=2 and $M$=6, with several time-frame measurements combined into a two-dimensional correlation image. }
		\label{fig1}
	\end{center}
\end{figure*}

We formulate and realize an application of the metasurface-based interferences for multi-photon quantum state measurement and reconstruction.
We develop a metasurface incorporating a set of $M/2$ interleaved metagratings (see \ref{sec:interleaving} in Supplementary Material), each composed of nano-resonators with specially varying dimensions and orientation according to the principle of geometric-phase~\cite{Bomzon:2001-1424:OL} to split specific elliptical polarization states~\cite{Mueller:2017-113901:PRL}, which would not be possible with conventional gratings (see \ref{sec:of} and \ref{sec:principle} in Supplementary Material).
This performs quantum projections in a multi-photon Hilbert space to $M$ imaging spots, each corresponding to a different elliptical polarization state [Fig.~\ref{fig1}(A)], which is essential
to minimize the error amplification in quantum state reconstruction~\cite{Foreman:2015-263901:PRL}. 
Then, by directly measuring all possible $N$-photon correlations from the $M$ output beams, it becomes possible to reconstruct the initial $N$-photon density matrix providing full information on the multi-photon quantum entanglement.
For example, in Fig.~\ref{fig1}(B) we show a sketch of three gratings (top) which realize an optimal set of projective bases shown as vectors on the Poincar\'e sphere (middle) for $M=6$. 

\begin{figure*}
	\begin{center}
		\includegraphics[width=0.85\textwidth]{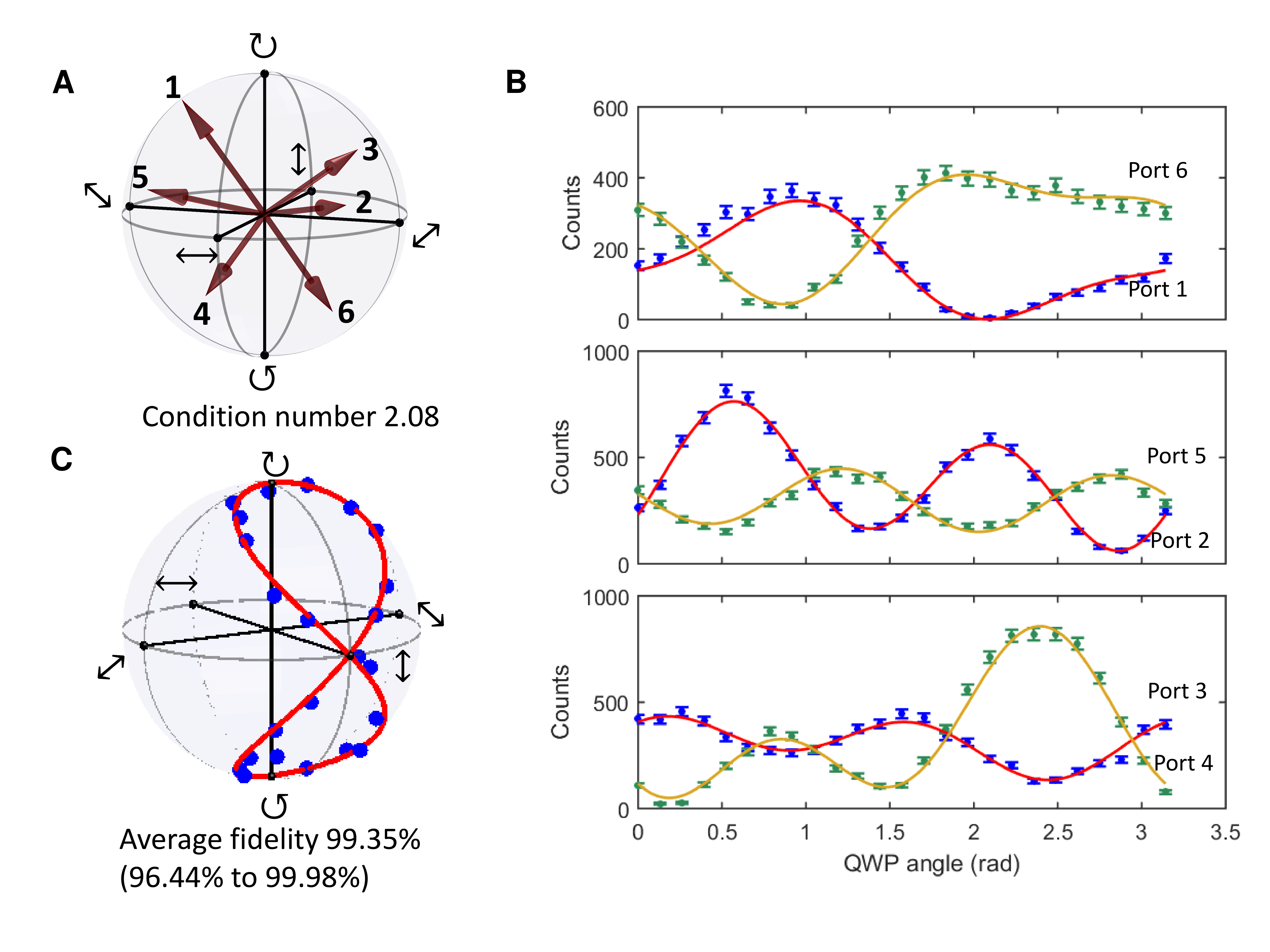}
		\vspace{-15pt}
		\caption{\textbf{Experimental measurement of heralded single-photon states with the metasurface.}  (A)~Classically characterized projective bases of the metasurface for ports numbered 1 to 6. (B)~Accumulated single-photon counts in each of $M$=6 output ports vs. the angle of a quarter-wave plate realizing a photon state transformation before the metasurface. Experimental data are shown with dots, with error bars indicating shot noise. Solid lines represent theoretical predictions based on classically measured metasurface transfer matrix.
(C) Comparison between the prepared (solid line) and reconstructed (dots) states based on the measurements presented in~(B), plotted on a Poincar\'e sphere.  }
		\label{fig2}
	\end{center}
\end{figure*}

The photon correlations between $M$ output ports can be obtained with simple polarization-insensitive click single-photon detectors.
The metasurface can be potentially combined with
single-photon sensitive electron-multiplying CCD (EMCCD) cameras~\cite{Edgar:2012-984:NCOM,Reichert:2018-7925:SRP} to determine the spatial correlations by processing multiple time-frame images of quantum states. 
We consider quantum states with a fixed photon number $N$, which is a widely-used approach in photon detection~\cite{James:2001-52312:PRA,Titchener:2016-4079:OL,Oren:2017-993:OPT,Titchener:2018-19:NPJQI}. 
The $N$-fold correlation data, stored in an array with $N$ dimensions, are obtained by averaging the coincidence events over multiple time frames.
For example, in Fig.~\ref{fig1}(C) we sketch a case with $N=2$ and $M=6$. In each frame, two photons arrive at different combinations of spots. After summing up the coincidence events over multiple time frames, we obtain a correlation in two-dimensional space. 
Following the general measurement theory of Ref.~\cite{Titchener:2018-19:NPJQI}, we establish that for an indistinguishable detection of $N$-photon polarization states (i.e. the detectors cannot distinguish which is which of the $N$ photons), the required number of output ports to perform the reconstruction scales linearly with the photon number as $M\geq N+3$, see Fig.~\ref{fig1}(B, bottom).
For instance, with $M=6$ up to $N=3$ photon states can be measured.

The parallel realization of multi-photon interferences with a single metasuraface offers practical advantages for quantum state measurements.
Conventional quantum state tomography~\cite{James:2001-52312:PRA} methods based on reconfigurable setups can require extra time and potentially suffer from errors associated with the movement of bulk optical components~\cite{James:2001-52312:PRA} or tuning of optical interference elements~\cite{Shadbolt:2012-45:NPHOT}.
Moreover, the conventionally-used sequential implementations of projective measurements present a fundamental limit for miniaturization, while being inherently sensitive to fluctuations or misalignment between different elements, especially for higher photon-number states.
The emerging 
methods based on static transformations implemented with bulk optical components~\cite{Bayraktar:2016-20105:PRA} or integrated waveguides~\cite{Titchener:2016-4079:OL,Oren:2017-993:OPT,Titchener:2018-19:NPJQI}
still require
multiple 
stages of interferences.
In contrast, our quantum metasurface provides an ultimately robust and compact solution,
the speed of which is only limited by the detectors.

We fabricate silicon-on-glass metasurfaces with $M=6$ and $M=8$ using standard semiconductor fabrication technology (see \ref{sec:fabMS} and \ref{sec:mschar} in Supplementary Material for details). 
The experimentally determined polarization projective bases obtained through classical characterization are plotted on the Poincar\'e sphere in Fig.~\ref{fig2}(A) for a metasurface with $M=6$ that is used later for quantum experiments. The transfer matrix measurements confirm that the polarization projective bases are close to the optimal frame.
The condition number, a measure of error amplification in the reconstruction (see \ref{sec:of} in Supplementary Material) is $2.08$, close to the fundamental theoretical minimum of $\sqrt{3} \simeq 1.73$. 
The reconstruction is immune to fabrication imperfections, as their effect is fully taken into consideration by performing an experimental metasurface characterization with classical light after the fabrication (see \ref{sec:imperfections} and \ref{sec:onSiteClassical} in Supplementary Material).

\begin{figure*}[t]
	\begin{center}
		\includegraphics[width=1\textwidth,height=0.69\textheight,keepaspectratio]{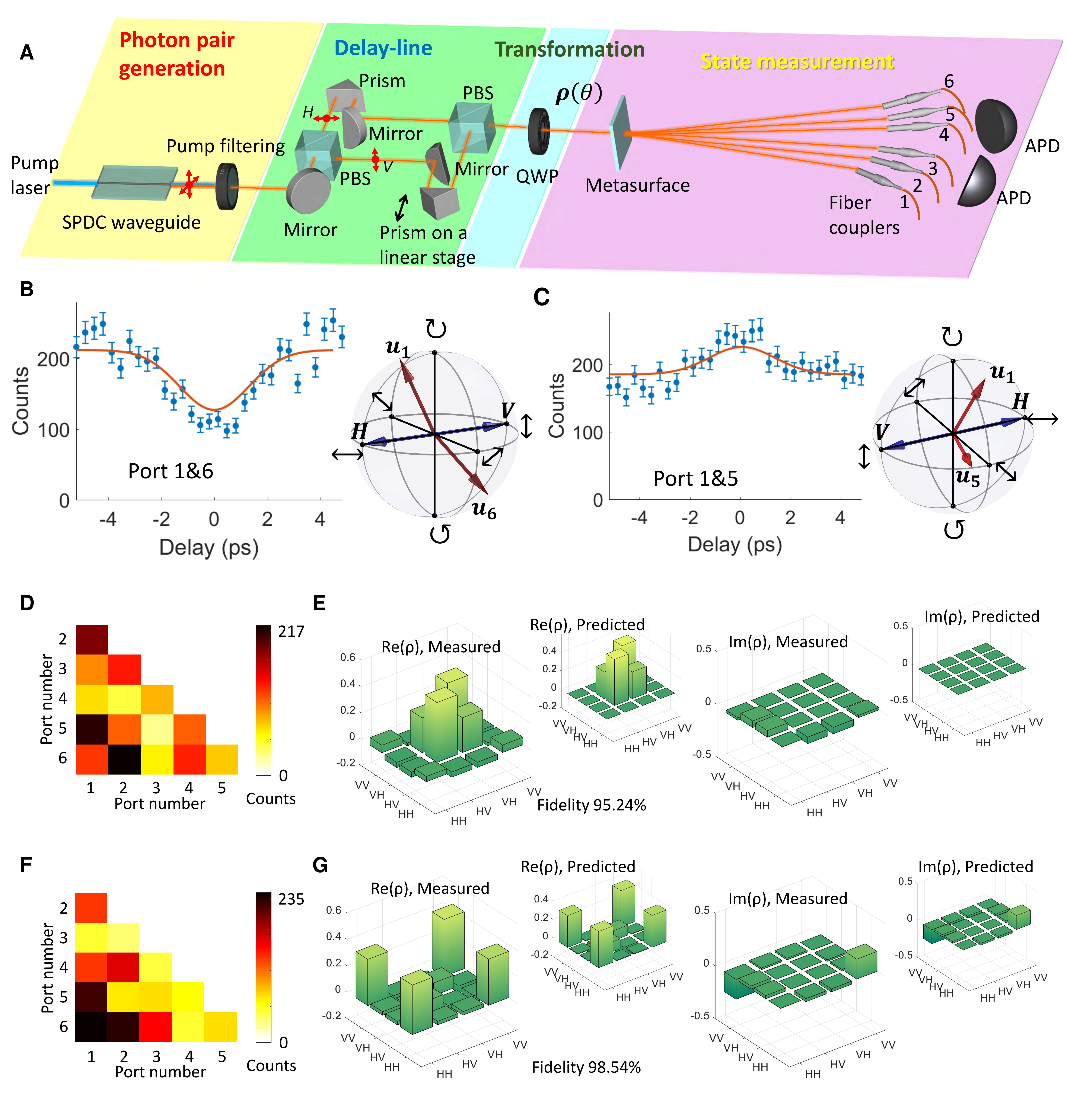}
		\vspace{-5pt}
		\caption{\textbf{Experimental two-photon interferences and state reconstruction with the metasurface.} (A) Schematic setup including photon pair generation and pump filtering, a delay-line with polarizing beam splitters (PBS) to control the path difference between orthogonally polarized photons in a pair, state transformation with a quarter-wave plate (QWP), and state measurement with the metasurface using avalanche photo-diodes (APDs). (B),(C)~Quantum correlations between ports (B)~1 and 6 with close-to-orthogonal bases and (C)~1 and 5 with non-orthogonal bases, shown with dots and error bars indicating shot noise. Solid curves represent theoretical predications. Red arrows in the Poincar\'e spheres denote projective bases of different ports. Blue arrows indicate the polarization state of entangled photons, with one photon in H- and the other in V-polarization. 
(D),(F)~Representative two-fold correlation measurements and (E),(G) the corresponding reconstructed density matrices $\rho$ labeled 'Measured' alongside with the theoretically predicted states labeled 'Predicted' for QWP orientations (D),(E)~$\theta=0^\circ$ and (F),(G)~$\theta=37.5^\circ$. }
		\label{fig3}
	\end{center}
\end{figure*}

First, we 
show that our metasurface enables accurate reconstruction of the quantum-polarization state of single photons.
A heralded photon source is used at a wavelength of $1570.6$~nm based on spontaneous parametric down conversion (SPDC) in a nonlinear waveguide (see \ref{sec:fabNL}, \ref{sec:SPDC}, \ref{sec:ppg}, and \ref{sec:expHeralded} in Supplementary Material for details). 
The heralded single photons are initially linearly polarized. They are  prepared in different polarization states by varying the angle of a quarter wave-plate (QWP), sent to the metasurface, and each diffracted photon beam is collected by a fiber-coupled interface to the single-photon detectors. 
By measuring the correlations with the master detector, we reconstruct the quantum-polarization state from the photon counts at the six ports. The results are shown in Fig.~\ref{fig2}(B), where
the curves are theoretical predictions and dots are experimental measurements. We observe that the measurement errors are dominated by the single-photon detection shot noise, which is proportional to the square root of the photon counts, as indicated by the error bars. 
We use the measured photon counts to reconstruct the input single-photon states by performing a maximum-likelihood estimation~\cite{James:2001-52312:PRA} and plot them on a Poincar\'e sphere in Fig.~\ref{fig2}(C). The reconstructed states present a high average fidelity of 99.35\% with respect to the prepared states.

Next, we realize two-photon interference, 
the setup of which is conceptually sketched in Fig.~\ref{fig3}(A). The SPDC source generates a photon pair with horizontal (H) and vertical (V) polarizations, with their path length difference controllable by a delay-line (see \ref{sec:expTwoPhoton} in Supplementary Material for details). We measure the effect of delay on the two-photon interference,
analogous to the Hong-Ou-Mandel (HOM) experiment~\cite{Hong:1987-2044:PRL}.
In such a nontrivially generalized two-photon interference, we expect a dip or peak depending on the $2\times 2$ transfer matrix $\mathbf{T_{ab}}\propto\left[\mathbf{u_a},\mathbf{u_b}\right]^\dagger$ from the two-dimensional polarization state vector to a chosen pair of ports, where $\dagger$ denotes transpose conjugate, and $\mathbf{u}_a$, $\mathbf{u}_b$ are the projective bases of ports $a$ and $b$, respectively. We note that $\mathbf{T_{ab}}$ corresponds to an effective Hermitian Hamiltonian resulting in a conventional HOM dip only if $\mathbf{u}_a$ and $\mathbf{u}_b$ are orthogonal, while otherwise a HOM peak can appear analogous to a lossy beam-splitter~\cite{Vest:2017-1373:SCI}.
Here we set the angle of the QWP at $\theta=0^\circ$, which means that the photon pairs are in a state $\mathbf{\rho} (\theta=0^\circ)$, where one photon is H- and another is V-polarized. 
As reflected in the Poincar\'e plot of Fig.~\ref{fig3}(B, right), where the red arrows denote projective bases of the two ports ($\mathbf{u}_a$, $\mathbf{u}_b$) and blue arrows represent the polarization of the photon pairs -- one photon in H- and the other in V-polarization, we see that the state vector $\mathbf{u_1}$ points to the opposite direction of $\mathbf{u_6}$. We find that in this case photons with cross-polarized entanglement in H-V basis will give rise to a dip in the interference pattern with the variation of path length difference, see Fig.~\ref{fig3}(B, left). Such a behavior is directly caused by the coalescence nature of bosons. The situation is quite different if we measure such an interference between ports $a=1$ and $b=5$, since $\mathbf{u_1}$ and $\mathbf{u_5}$ are far from being orthogonal. This can be seen from the red arrows in the Poincar\'e sphere of Fig.~\ref{fig3}(C, right), where the angle between the two vectors representing $\mathbf{u_1}$ and $\mathbf{u_5}$ is much smaller than $\pi$. For entangled photons with H and V polarization in a pair, interference under the transfer matrix $\mathbf{T_{15}}$ leads to a peak instead of a dip when varying the path difference in the delay-line. Indeed, in Fig.~\ref{fig3}(C, left) we observe a peak, which is related to the anti-coalescence of bosons in transformations induced by non-Hermitian Hamiltonians, a nontrivial generalization of the HOM interference analogous to Ref.~\cite{Vest:2017-1373:SCI}. For details of the theoretical predictions and experimental methods see \ref{sec:fabMS} in Supplementary Material.

As a following step, we measure all 15 two-fold nonlocal correlations between the $M=6$ outputs from the metasurface for a given input state where the time delay is fixed to zero. This provides us full information to accurately reconstruct the input two-photon density matrix. 
We use two single-photon detectors to map out all possible output combinations, while this could be potentially accomplished even simpler with an EMCCD camera.
We show representative results for two different states $\mathbf{\rho}(\theta=0^\circ)$ and $\mathbf{\rho}(\theta=37.5^\circ)$ in Figs.~\ref{fig3}(D,E) and \ref{fig3}(F,G), respectively. Note that $\mathbf{\rho}(\theta=0^\circ)$ is a state where photon pairs have cross-polarized entanglement beyond the classical limit, yet it is not fully pure (see \ref{sec:fabMS} in Supplementary Material), providing a suitable test case for reconstruction of general mixed states. 
In Fig.~\ref{fig3}(D) we show the measured two-fold correlations for the input state $\mathbf{\rho}(\theta=0^\circ)$, and the reconstructed density matrix is shown in Fig.~\ref{fig3}(E). The fact that only the bunched four central elements are non-zero confirms cross-polarized property of our photon pairs in H-V basis. Moreover, the non-zero $|{HVVH}\rangle$ element implies the presence of two-photon entanglement. It is smaller compared to the diagonal element $|{HVHV}\rangle$, indicating that the polarization state is not fully pure. While $\mathbf{\rho}(\theta=0^\circ)$ only has non-zero elements in the real part of the density matrix, 
we also show the measurement and reconstruction of $\mathbf{\rho}(\theta=37.5^\circ)$ that contains nontrivial imaginary elements in Figs.~\ref{fig3}(F,G). In both cases, we achieve a very good agreement between the predicted and reconstructed density matrices as evidenced by high fidelity exceeding 95\%. The correlation counts are obtained by a Gaussian fitting to the correlation histogram to remove the background, which is less than 10\% of the signal for all measurements shown in Fig.~\ref{fig3}(F), see details in \ref{sec:expTwoPhoton} of Supplementary Material.

Our results illustrate the manifestation of multi-photon quantum interference on metasurfaces. We formulate a concept of parallel quantum state transformation with metasurfaces, enabling single- and multi-photon state measurements solely based on the interaction of light with sub-wavelength thin nanostructures and nonlocal correlation measurements without a requirement of photon-number-resolvable detectors. This presents the ultimate miniaturization and stability combined with high accuracy and robustness, as we demonstrate experimentally via reconstruction of one- and two-photon quantum-polarization states including the amplitude, phase, coherence and quantum entanglement. In general, our approach is particularly suitable for imaging-based measurements of multi-photon polarization states, where the metasurface can act as a quantum lens to transform the photons to a suitable format for the camera to recognize and retrieve more information. Furthermore, there is a potential to capture other degrees of freedom associated with spatially varying polarization states for the manipulation and measurement of high-dimensional quantum states of light, with applications including free-space communications and quantum imaging.

\begin{center}
{\bf Acknowledgment}
\end{center}

We gratefully thank Hans Bachor, Marlan Scully, Ian Walmsley and Frank Setzpfandt for fruitful discussions, Roland Schiek and Yair Zarate for help in developing ovens for waveguide temperature control, and Mingkai Liu for advice on numerical simulations. 
\textbf{Funding:} This work was supported by the Australian Research Council (including projects DP160100619, DP150103733, DE180100070); the Ministry of Science and Technology (MOST), Taiwan under contracts 106-2221-E-008-068-MY3.
A portion of this research was conducted at the Center for Nanophase Materials Sciences, which is a DOE Office of Science User Facility. 
\textbf{Author contributions:} K.W., D.N.N., and A.A.S. conceived and designed the research; K.W. and L.X. performed numerical modeling of metasurface design; S.S.K. and I.I.K. fabricated the dielectric metasurfaces; H.P.C. and Y.H.C. fabricated nonlinear waveguides; K.W., J.G.T., H.P.C., M.P., and A.S.S. performed optical experimental measurements and data analysis; A.A.S, D.N.N., and Y.S.K. supervised the work; K.W., A.A.S, D.N.N., and Y.S.K. prepared the manuscript and supplementary in contact with all authors.

\clearpage
\newpage

\newpage
\begin{center}
{\Large Supplementary Material}
\end{center}
\renewcommand{\thefigure}{S$\the\numexpr\value{figure}-3$}

\tableofcontents

\section{Choice of projective bases for optimal-frame polarization measurements}\label{sec:of} 

\begin{figure*}[t]
	\centering
	\includegraphics[width=0.68\textwidth]{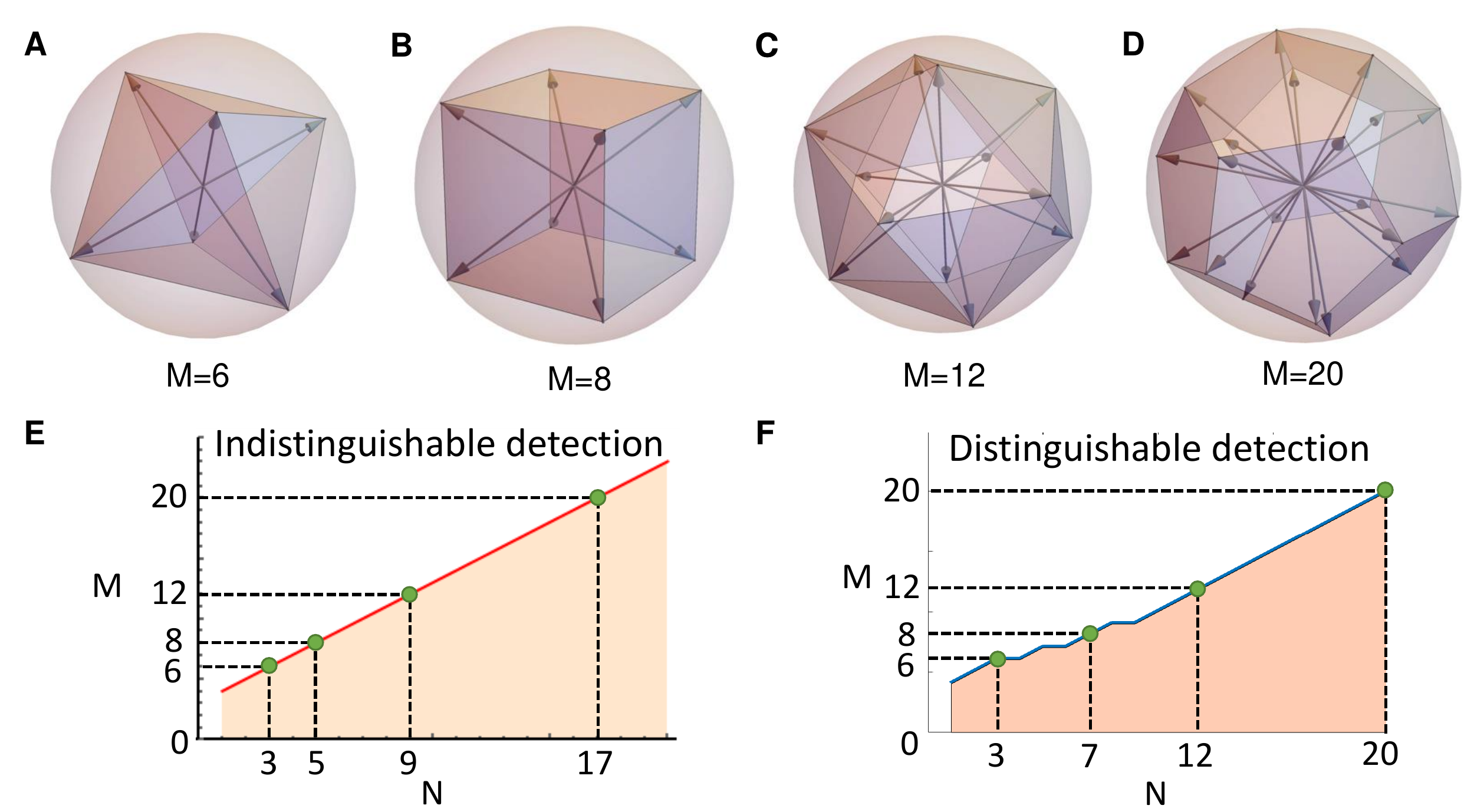}
	\caption{(A--D) Platonic polyhedra for number of vertices 6, 8, 12, 20 embedded in a sphere. The arrows constitute a set of optimal-frame polarization state reconstruction bases on a Poincar\'e sphere. (E,F) Minimum number of ports $M$ required for the reconstruction of states with the photon number $N$ for different detection schemes: (E) the single-photon detectors do not distinguish specific photons among the $N$ photons; (F) the photons are distinguishable by the detectors. 
}\label{fig:sca}
\end{figure*}

The so-called optimal frame consideration of choosing projective bases depends mainly on two factors: (i) inverse condition of the transfer matrix; (ii) for Fock states $\ket{N}$ ($N$-photon number state), possibility to reconstruct states with maximum $N$ using on-off click detectors. The first factor can be easily understood in a sense that the transfer function of the metasurface (from elements in its density matrix to the $N$-fold correlations between spatially diffracted spots) should provide a reversible relation between the output and the input, to enable a ``well-conditioned'' reconstruction. This can also be intuitively understood by spreading out the projective basis vectors evenly on the Poincar\'e sphere, such that they can fully probe any unknown state. Depending on the number of outputs $M$, there are different choices (see Ref.~\cite{Foreman:2015-263901:PRL}). Actually there is a third condition: (iii) one does not want to lose photons (i.e. no polarizers or other lossy polarizing optics) to efficiently use the photons, which can be satisfied when the bases are composed of pairs of orthogonal polarizations. Choices fulfilling the above three conditions can be obtained with platonic polyhedra in spherical $t$-design, with $M=6,8,12,20,\cdots$~\cite{Foreman:2015-263901:PRL}. For example, in Fig.~\ref{fig:sca}(A--D) we show such platonic solids and the projective bases (arrows) in a sphere for $M=6,8,12,20$, respectively. Note that the set of bases have a rotational degree of freedom without changing the optimal-frame nature.  

For classical polarization measurements, one typically chooses the minimum three pairs of projection bases [$M=6$, see Fig.~\ref{fig:sca}(A)]. Since the platonic solid has a rotational degree of freedom, it is convenient to choose horizontal, vertical, diagonal, anti-diagonal linear polarizations and left-hand, right-hand circular polarizations. However, the second factor above determines that for multi-photon quantum measurements, a larger output number $M$ is necessary if the detectors/cameras cannot resolve photon number. The minimum number of output ports $M$ depending on the photon number $N$ is shown in Fig.~\ref{fig:sca}(E,F), for a more general theory see Ref.~\cite{Titchener:2018-19:NPJQI}. In our experiment, we use the so-called indistinguishable detection scheme, where the detector does not know which photon is which. In this case, there are $M!/[N!(M-N)!]$ independent correlation elements out of the $M$ ports. 
The minimum number of ports $M$ depends linearly on $N$, i.e. $M\geq N+3$. For $M=6, 8, 12, 20$, the maximum $N$ of states that can be reconstructed is $N=3,5,9,17$, respectively [see Fig.~\ref{fig:sca}~(E)]. There also exists another case, where the detection can distinguish the photons, such as by using another degree of freedom than polarization. For instance, if $N$ photons are in $N$ different paths, then one can use $N$ copies of the same metasurfaces to perform the measurement, placing one metasurface in each path. Another example is when $N$ photons can be distinguishable in frequency, then one can use detectors or camera pixels embedded with different frequency filters to distinguish them. In this case, the dependence of minimum output number $M$ on $N$ becomes more complicated, as given by $M!/(M-N)!\geq 2^{2N}$, which is shown in Fig.~\ref{fig:sca}~(F). We mark the cases for $M=6, 8, 12, 20$, where the maximum $N$ for states that are possible to reconstruct is $N=3,7,12,20$, respectively.

A widely used and convenient measure of the reversibility of the transfer matrix is the so-called condition number. Inherently, condition number quantifies the amplification of error (standard deviation) in the reconstruction. Quite generally, consider a vector $\mathbf{S}$ transformed to vector $\mathbf{V}$ after a linear transformation described by the matrix $\mathbf{A}$:
\begin{equation}
\mathbf{V}=\mathbf{A}\mathbf{S}.
\end{equation}
Specifically, $\mathbf{S}$ can be the Stokes vector of classical light ($4\times1$) to be measured. We note that for a single-photon polarization state ($N=1$), the decomposition of its density matrix in the Pauli matrices and the identity matrix can form a mathematically equivalent vector as the classical $4\times1$ Stokes vector. $\mathbf{A}$ is the transfer matrix ($M\times 4$) of the metasurface with each row denoting a Stokes vector that the state is projected to, and $\mathbf{V}$ is a vector representing the observables from the $M$ outputs ($M\times1$). The observable for classical light is power, whereas for single-photon states it will be the probabilities of photons getting out from the output ports of the system.
In the reconstructed $\mathbf{S}$ (denoted as $\mathbf{S_r}$) using $\mathbf{V}$, one has to use the inverse (or pseudo-inverse) $\mathbf{A}^{-1}$:
\begin{equation} \label{eq:rec}
\mathbf{S_r}=\mathbf{A}^{-1}\mathbf{V}.
\end{equation}
Now consider an error vector $\Delta \mathbf{V}$ in the measurement, which will then propagate to another error vector $\Delta \mathbf{S}$ in the reconstructed $\mathbf{S_r}$:
\begin{equation}
\Delta \mathbf{S}=\mathbf{A}^{-1}\Delta \mathbf{V}.
\end{equation}
A direct measure of the error is its norm (essentially the standard deviation), denoted by $||\cdot||$. According to the Cauchy's Inequality, $||\Delta \mathbf{S}||$ has an upper bound
\begin{equation}
||\Delta \mathbf{S}||\leq ||\mathbf{A}^{-1}||\ ||\Delta \mathbf{V}||.
\end{equation}
For the relative error with respect to the reconstructed vector $||\Delta \mathbf{S}||/||\mathbf{S_r}||$,
using Eq. \eqref{eq:rec} and again the Cauchy's Inequality we can find its upper bound
\begin{equation}
\frac{||\Delta \mathbf{S}||}{||\mathbf{S_r}||} \leq ||\mathbf{A}||\ ||\mathbf{A}^{-1}||  \frac{||\Delta \mathbf{V}||}{||\mathbf{V}||},
\end{equation}
where $||\mathbf{A}||\ ||\mathbf{A}^{-1}||$ can act as a measure of the amplification of the relative error in such a system. In other words, it quantifies the robustness of the system against error propagation. Therefore such a quantity is defined as the condition number
\begin{equation}
\kappa(\mathbf{A})=||\mathbf{A}||\ ||\mathbf{A}^{-1}||,
\end{equation}
which is a measure of the inverse condition in reconstruction problems. The fundamental limit for the condition number in classical or single-photon Stokes vector reconstruction is $\sqrt{3}$ if $||\cdot||$ is the Euclidean norm~\cite{Foreman:2015-263901:PRL}, which appears with the optimal-frame choice of projective bases. As discussed, elements in a single-photon Stokes vector are basically the decomposition of a single-photon density matrix $\mathbf{\rho}$ into the identity matrix $\mathbf{I}$ and the three Pauli matrices $\mathbf{\sigma_x}, \mathbf{\sigma_y}, \mathbf{\sigma_z}$, i.e. 
\begin{equation}
\mathbf{\rho}=S_0 \mathbf{I} + S_1 \mathbf{\sigma_z} + S_2 \mathbf{\sigma_x} + S_3 \mathbf{\sigma_y}.
\end{equation}
Similarly, a multi-photon density matrix ($N$-photons) can also be decomposed to a set of multi-photon Pauli matrices by tracing out all possible tensor product of $N$ matrices out of $\mathbf{I}$ and $\mathbf{\sigma_x}, \mathbf{\sigma_y}, \mathbf{\sigma_z}$. For such a multi-photon Stokes vector $\mathbf{S}_N$, the transfer matrix is expressed as $N$ tensor products of $\mathbf{A}$:
\begin{equation}
\mathbf{V}_N=\mathbf{A}^{\otimes N}\mathbf{S}_N,
\end{equation}
where $\mathbf{V}_N$ becomes an $MN$-dimensional vector as one traces out all the $N$-fold correlations out of the $M$ outputs. Using the lower limit of the condition number for single-photon transfer matrix, we can also obtain the achievable minimum condition number for $N$-photon measurements:
\begin{equation}
\kappa_N\geq \left(\sqrt{3}\right)^N.
\end{equation}

From the above analysis one can see that unlike classical polarization measurements, if one targets higher photon number states where photon number is also a building block of the space, it is vital to use larger $M$. Then, different from those metasurfaces designed for classical measurements, where they can also project states to linear and circular polarizations, our metasurface needs to enable the projection of the states to arbitrarily chosen pairs of polarization states, including elliptical polarizations. Note that despite for $M=6$ typically one uses four linear polarizations and two circular polarizations as the projective bases, here in our experiment we use all  elliptical polarization bases to test the capability of the quantum metasurface, as for $M>6$, elliptical polarization basis will be necessary. 

\section{Working principles of the metagratings}  \label{sec:principle} 

The metasurface is assembled from $M/2$ metagratings, with each one diffracting a pair of specially chosen polarization states to two directions symmetric with respect to the transmissive axis. 
The novel aspect of our geometric-phase metagrating lies in the feature that we extend the capability to project our states to any pairs of elliptical polarization states. As discussed above in \ref{sec:of}, this is a crucial property for quantum-polarization measurements -- unlike classical polarization measurement, for multi-photon quantum states more diffracted polarization components are needed for on-off click detectors. This has been considered difficult to realize with non-chiral birefringent structures, until recently a generalized geometric phase was implemented in Ref.~\cite{Mueller:2017-113901:PRL} for polarization sensitive hologram. Here we provide the theory for designing metagratings that can achieve decomposition of elliptical polarization components. 

\begin{figure}[t]
	\centering
	\includegraphics[width=\columnwidth]{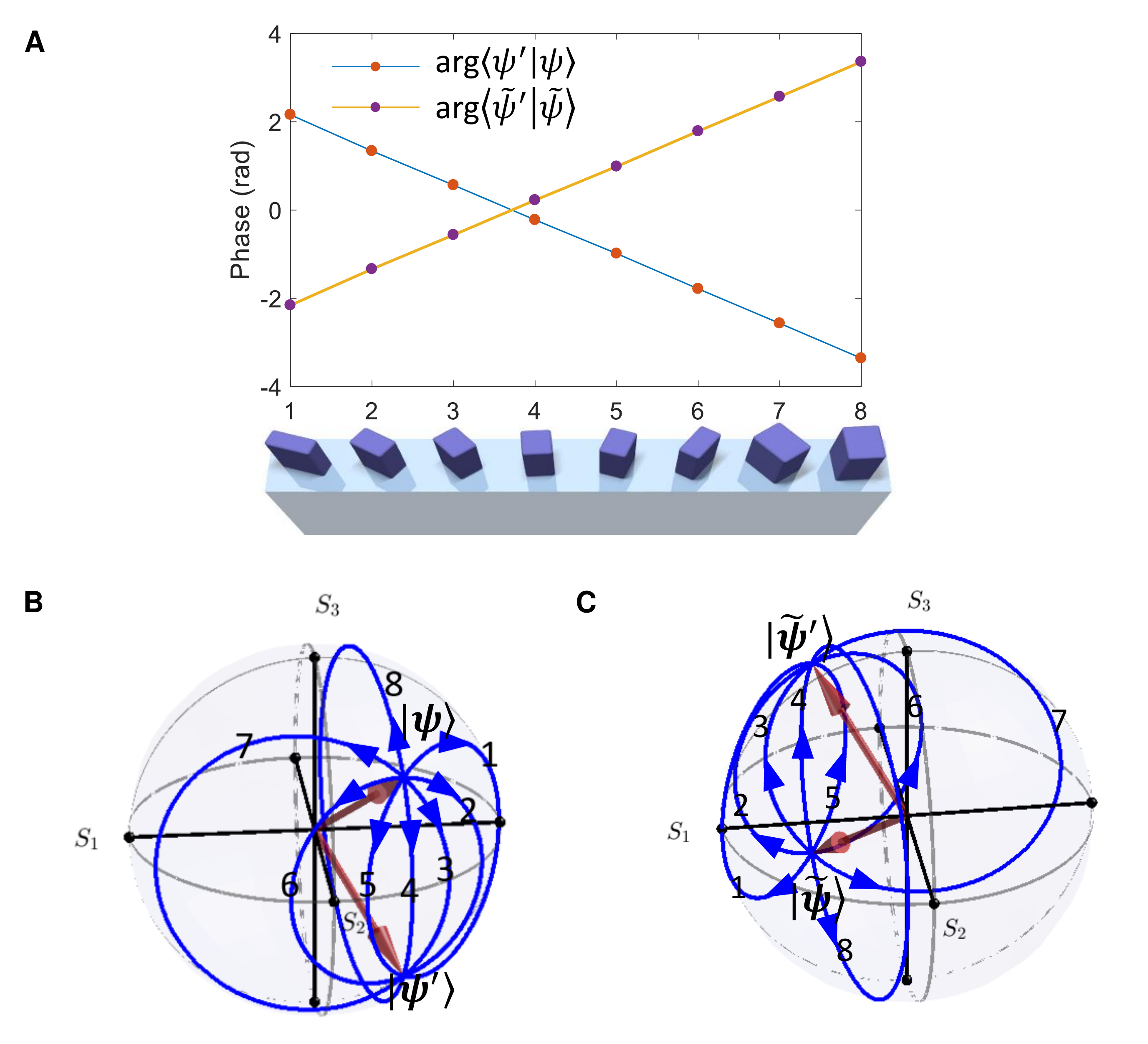}
	\caption{Working principle of a metagrating that spatially diffracts a pair of elliptical polarization components to two directions, where a metagrating with a super-cell of 8 meta-atoms is shown: (A) Conceptual 3D structure of the metagrating with periodic boundaries extending to the left and right sides, with the phases picked up by $\ket{\psi}$ and $\ket{\tilde{\psi}}$ shown as dots for each meta-atom; (B) The paths that $\ket{\psi}$ goes on the Poincar\'e sphere, with each meta-atom changing the handedness of $\ket{\psi}$ by going along paths at different angles; (C) Similar paths for $\ket{\tilde{\psi}}$. 
}\label{fig:grat}
\end{figure}

We assume that each meta-atom behaves as a birefringent crystal with its fast and slow axes in the plane of the metasurface, depending on the orientation angle $\theta$ of its fast axis. We denote the phase picked up along the fast and slow axes as $\phi_1$ and $\phi_2$, respectively ($\phi_1<\phi_2$). A single-photon pure state can be described by the state vector $\boldsymbol{\psi}_\mathrm{in}=\left[c_\mathrm{H},c_\mathrm{V} \right]^\mathrm{T}$, where $c_\mathrm{H}$ and $c_\mathrm{V}$ are the wavefunctions of photons polarized in horizontal (x) and vertical (y) directions, respectively. Here $^\mathrm{T}$ represents the matrix transpose. The transformation that each meta-atom does can be expressed by a $2\times 2$ matrix $\mathbf{U}$
\begin{equation}\label{eq:jonesmatrix}
\mathbf{U}
=
\left[
\begin{array}{ccc}
\cos \theta & -\sin \theta \\
\sin \theta & \cos \theta
\end{array}
\right]
\left[
\begin{array}{ccc}
e^{i\phi_1} & 0 \\
0 & e^{i\phi_2}
\end{array}
\right]
\left[
\begin{array}{ccc}
\cos \theta & \sin \theta \\
-\sin \theta & \cos \theta
\end{array}
\right].
\end{equation}
The eigenstates of the static Hamiltonian that generates $\mathbf{U}$ are always pairs of orthogonally-polarized \emph{linear} polarization states. The angle $\theta$ determines the orientation of the two eigenstates. Such a fact means that $\mathbf{U}$ has two basic properties: (1) it transforms an orthogonal pair of polarization states to another orthogonal pair, which is governed by its Hermiticity; (2) such a transformation is not chiral, as the eigenstates are linear polarizations. With such properties, it can be easily found that for any pair of elliptical polarizations, changing their handedness can be realized with any $\theta$, as long as one chooses proper difference between $\phi_1$ and $\phi_2$. The phase we utilize is directly related to the Pancharatnam-Berry phase. The working principle is conceptually shown in Fig.~~\ref{fig:grat}(A). The meta-atoms have both varying orientation and $\phi_1$, $\phi_2$. They are tailored in a way that each meta-atom changes the handedness of a pair of elliptical polarization states $\ket{\psi}$ and $\ket{\tilde{\psi}}$, changing them to $\ket{\psi'}$ and $\ket{\tilde{\psi'}}$, respectively. In such a transformation, each state acquires a linear phase gradient in space with inverse signs for $\ket{\psi}$ and $\ket{\tilde{\psi}}$.   

\begin{figure*}[t]
	\centering
	\includegraphics[width=0.75\textwidth]{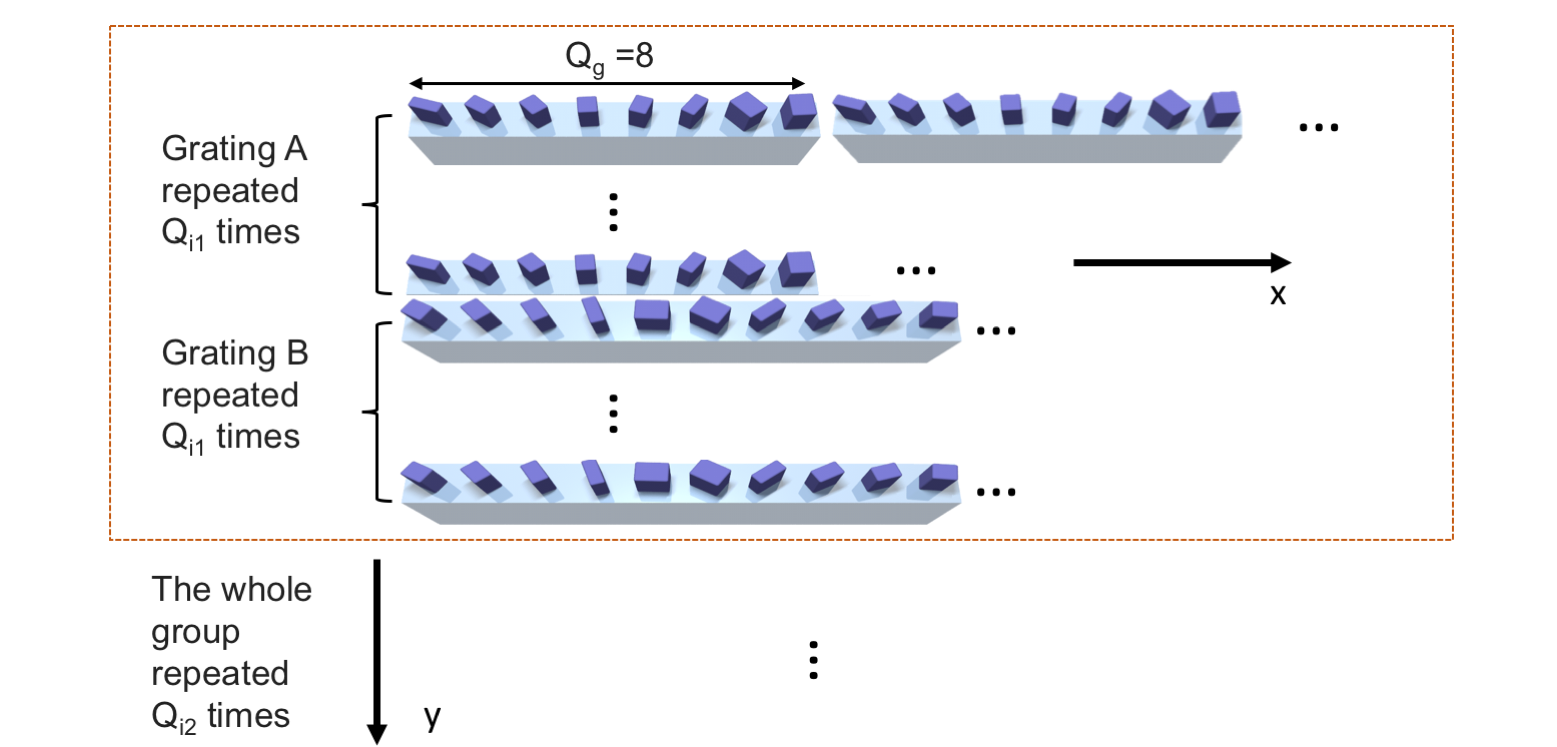}
	\caption{Schematic layout of interleaved metagratings on the metasurface, where two different gratings (named A and B) composed of 8 and 10 meta-atoms are shown as an example. 
}\label{fig:interleave}
\end{figure*}

For a more detailed description, we firstly write down $\ket{\psi}$ and $\ket{\tilde{\psi}}$:
\begin{equation}
\begin{split}
\ket{\psi}&=\left[\cos\ \alpha,\exp (i\beta)\ \sin\ \alpha\right]^\mathrm{T},\\
\ket{\tilde{\psi}}&=\left[-\sin\ \alpha,\exp (i\beta)\ \cos\ \alpha\right]^\mathrm{T}.
\end{split}
\end{equation}
Note that $\braket{\tilde{\psi}|\psi}=0$, showing the orthogonality of the two polarization states. As described above, we target changing the handedness of $\ket{\psi}$ and $\ket{\tilde{\psi}}$. Hence we are looking for the different $\mathbf{U} (\theta)$ that can achieve such a transformation:
\begin{equation}\label{eq:1}
\exp(i\gamma)\ket{\psi'}=\mathbf{U}(\theta)\ket{\psi},
\end{equation}
where $\ket{\psi'}=\left[\cos\ \alpha,\exp (-i\beta)\ \sin\ \alpha\right]^\mathrm{T}$. This can be seen from Fig.~\ref{fig:grat}(B) where $\ket{\psi}$ and $\ket{\psi'}$ are represented as two arrows on the Poincar\'e sphere, and different angles $\theta$ for different meta-atoms lead to different paths in converting the state $\ket{\psi}$ to $\ket{\psi'}$ with a reversed handedness. Similarly, after reversing the handedness of the orthogonal counterpart $\ket{\tilde{\psi}}$, it becomes $\ket{\tilde{\psi'}}=\left[-\sin\ \alpha,\exp (-i\beta)\ \cos\ \alpha\right]^\mathrm{T}$. So firstly we look for the solution of Eq.~\eqref{eq:1}. The general condition is
\begin{equation}\label{eq:2a}
|\bra{\psi'}\mathbf{U}(\theta)\ket{\psi}|=1.
\end{equation}
We numerically solve Eq.~\eqref{eq:1} at different angles $\theta$ to obtain $\phi_1 (\theta)$ and $\phi_2 (\theta)$. Note that the solution of $\phi_1 (\theta)$ and $\phi_2 (\theta)$ is not unique, however, we have conveniently taken $\phi_1=-\phi_2$ and hence simplified the calculation. Then, the phase picked up when converting $\ket{\psi}$ to $\ket{\psi'}$ can be easily calculated by
\begin{equation}\label{eq:2b}
\gamma(\theta)=\arg \bra{\psi'}\mathbf{U}(\theta)\ket{\psi}.
\end{equation}
When designing the metagrating, we introduce a spatially varying $\theta(x_n)$ for $x_n=(km+n)d$ where $k$ is an integer, $m$ is the number of meta-atoms (essentially the periodicity) and $d$ is the lattice constant for the meta-atoms. We design the metagrating such that 

\begin{equation}\label{eq:3a}
\gamma (n)=-\frac{2\pi}{m} n+\mathrm{C_1},
\end{equation}
where $C_1$ is a constant. It can be demonstrated that for the orthogonal counterpart $\ket{\tilde{\psi}}$, when its handedness is reversed and as it becomes $\ket{\tilde{\psi'}}$ [see Fig.~\ref{fig:grat}(C)], the phase $\gamma'$ it picks up is
\begin{equation}\label{eq:3b}
\gamma (n)=\frac{2\pi}{m} n+\mathrm{C_1},
\end{equation}
which has an inverse slope as $\ket{\psi}$. Therefore $\ket{\psi}$ and $\ket{\tilde{\psi}}$ get spitted spatially with inverse transverse state vectors. If the sampling rate is high enough (i.e. high enough number of meta-atoms in one supercell of the metagrating) and the phase slopes as shown in Fig.~\ref{fig:grat}(A) are really linear, the diffraction efficiency can be close to unity. 
Note that the typically used geometric-phase grating that splits two circular polarizations is a special case of the scheme given here. 

\begin{figure*}[t]
	\centering
	\includegraphics[width=0.9\textwidth]{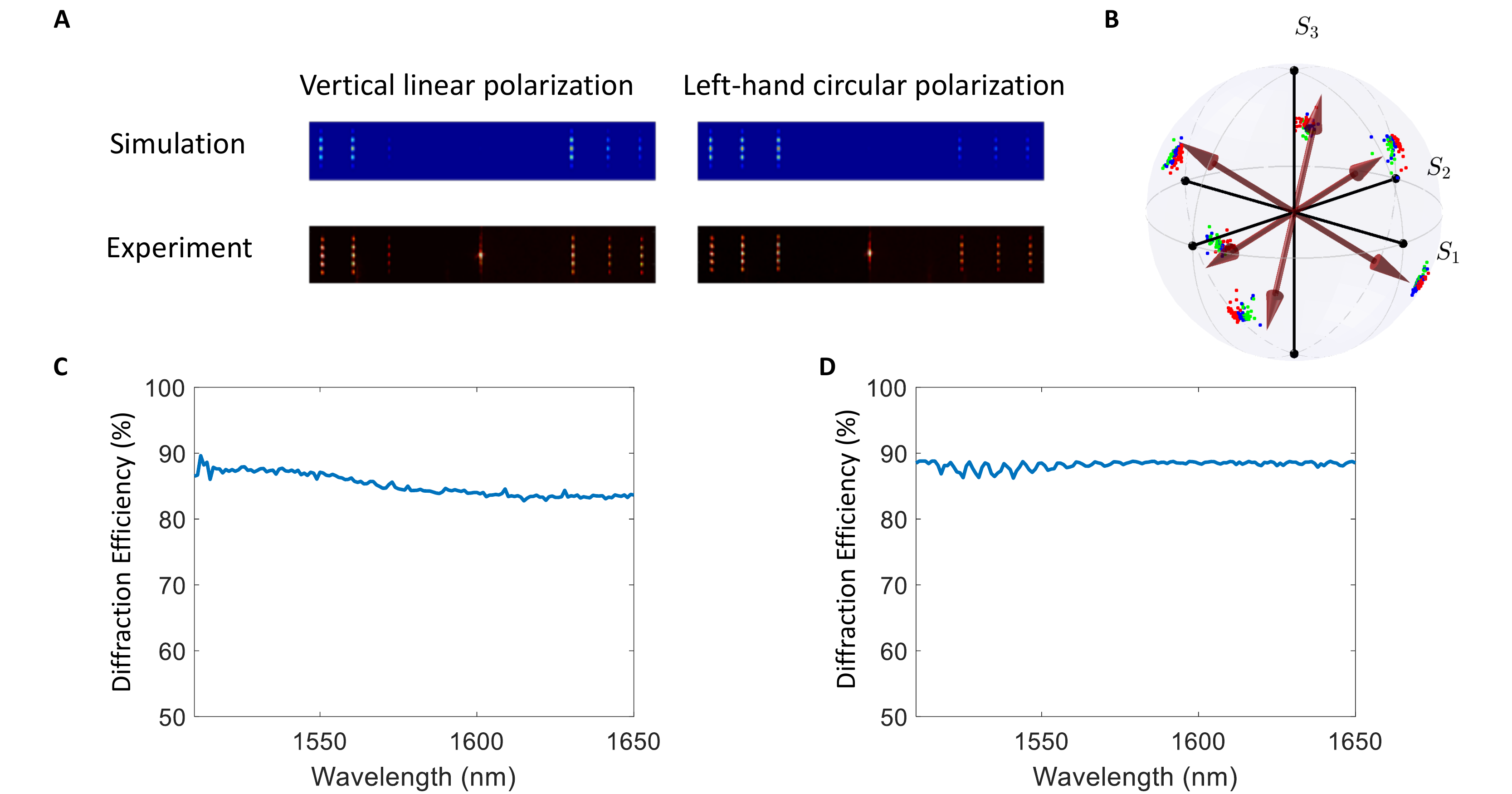}
	\caption{Classical characterization results of representative metasurfaces. (A) Far-field images of a 6-output metasurface for two different input polarizations, with comparison to simulation. (B) Characterized projection bases for the 6-output metasurface with colors denoting different wavelength ranges (red 1556--1585 nm, blue 1586--1615 nm, and green 1616--1650 nm). The arrows indicate the theoretically optimal design. (C,D) Diffraction efficiency, defined as the light power of the useful diffracted spots divided by the total power of all spots, measured across a broad band with the (C) 6-output and (D) 8-output metasurfaces. 
}\label{fig:classical}
\end{figure*}

\begin{figure*}
	\centering
	\includegraphics[width=.8\textwidth]{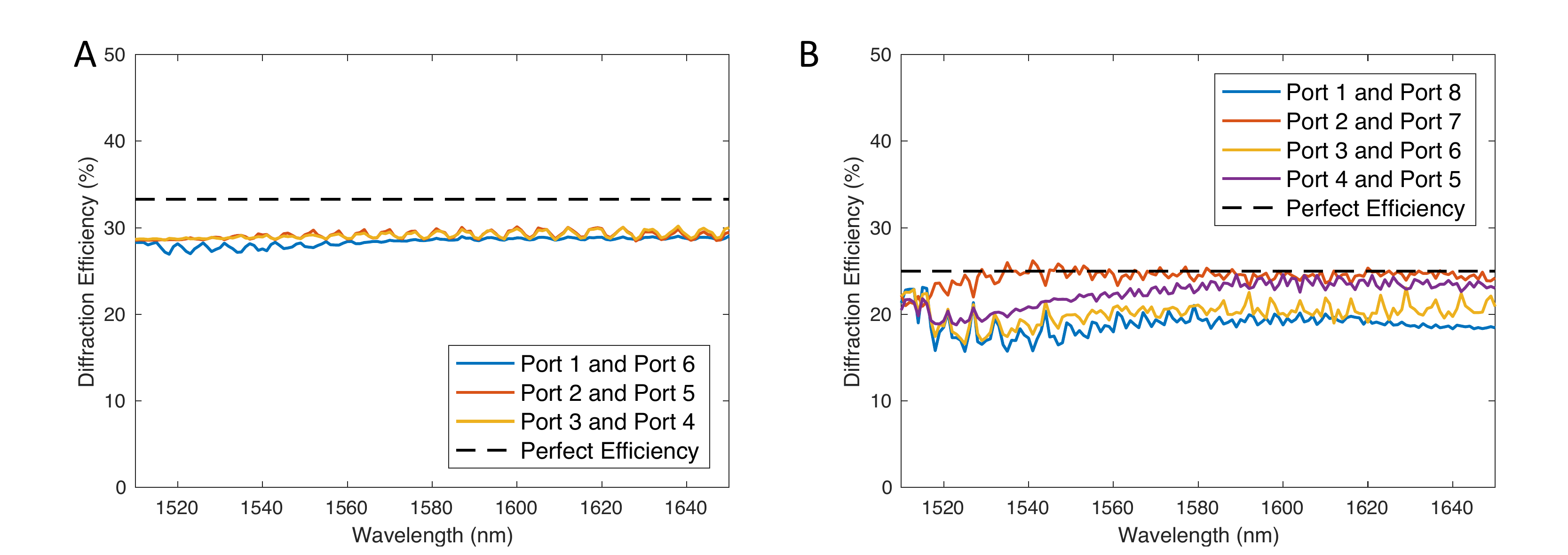}
	\caption{Diffraction efficiencies for pairs of ports associated with different interleaved gratings from the fabricated metasurfaces with (A) $M/2=3$ interleaved gratings and 6 ports and (B) $M/2=4$ gratings and 8 ports. Dashed lines indicate the theoretically optimal regime of (2/M).
}\label{fig:porteff}
\end{figure*}

\section{Arrangement of metagratings on the metasurface}  \label{sec:interleaving} 

The quantum metasurface consists of multiple metagratings, interleaved to generate the functionality of spreading out $M$ output polarization components in parallel. Here we briefly show how we interleave these metagratings and estimate the maximum number of different gratings one can incorporate in a metasurface.

Along the grating direction $x$ (see Fig.~\ref{fig:interleave}), the number of output ports which can be formed without overlapping can be estimated as follows. First, a grating period should consist of at least $Q_g$ meta-atoms, to perform reliable splitting of elliptical polarizations. Such a grating imposes the transverse wavevector component of $k_{g,max} = 2 \pi / (Q_g \Delta r)$, where $\Delta r$ is the spacing between neighboring meta-atoms (or, lattice constant). Second, the width of diffraction peaks from the gratings is defined by the metasurface size as $\Delta k = 2 \pi / L_x$. Then, the maximum number of gratings along the $x$ direction is $M/2 < k_{g,max} / \Delta k = L_x / (Q_g \Delta r)$.
In our metasurface, $L_x$ = 2mm, the minimum $Q_g=8$ (see Fig.~\ref{fig:interleave}), $\Delta r = 800 \mathrm{nm}$, and the maximum number of gratings is $M/2 < 312$.
 
In the vertical ($y$) direction as shown in Fig.~\ref{fig:interleave}, it is preferential to interleave the gratings to (i) lower down the diffracted transverse wave vector along $y$ to enhance the diffraction efficiency and (ii) minimize the dependence of the transfer matrix on the spatial beam profile. For (i), we repeat the same grating $Q_{i1}$ times before putting a different one, as illustrated in Fig.~\ref{fig:interleave}. For (ii), we ensure that the size along $y$ is large enough to repeat the whole group of $Q_{i1} \times M/2$ gratings several times ($Q_{i2}$), as shown in Fig.~\ref{fig:interleave}. Then, the maximum number of gratings along the $y$ direction is $
M/2 < L_y / ( Q_{i1} Q_{i2}  \Delta r).
$
Since $Q_g\ll Q_{i1} Q_{i2}$ for a square metasurface with $L_x \simeq L_y$, the main limitation comes from the grating interleaving in the vertical direction. In our metasurface design, $L_y = 2$ mm, $Q_{i1} = 100$, $Q_{i2} = 6$, $\Delta r = 800$ nm, such that $M/2 \leq 4$. By increasing the interleaving density to still practically suitable regime with $Q_{i1} = 50$, we could increase the maximum number of gratings interleaved in the $y$-direction to $M/2 \leq 8$, enabling the measurement of up to $N=13$ photon-number states. By also increasing the metasurface dimension to the experimentally feasible size of $L_y=5$ mm, up to $M/2 \leq 20$ gratings can be interleaved. There appears an interesting open research question on optimizing the metasurface design for encoding most efficiently a large number of gratings~\cite{Maguid:2017-e17027:LSA} for multiple quantum measurements.
 
Overall, the number of interleaved gratings $M/2$ scales linearly with the metasurface dimensions, which in combination with linear scaling vs. the number of photons [see Fig.~\ref{fig1}(B, bottom)] makes the platform suitable for characterization of multi-photon states.

\section{Fabrication of dielectric metasurfaces} \label{sec:fabMS} 

An 792-nm-thick poly-crystalline silicon (Poly-Si) thin film is prepared on a four-inch quartz wafer (thickness 500 $\mathrm{\mu m}$) via low-pressure chemical vapor deposition (LPCVD) in horizontal tube furnace. A layer of Polymethyl Methacrylate (950K PMMA A4) covers the sample surface as a positive tone resist via a spinner. Then the sample is exposed to electron beam lithography (EBL, model JEOL 9300FS) to obtain the pattern of the metasurfaces. Since silicon is only semi-conductive, before the EBL a 10 nm layer of chromium (Cr) is evaporated on top of the photo-resist for charge dissipation. After the EBL, wet chemistry removal of the Cr layer and sequential development are performed. Then, another 20-nm-thick Cr layer is evaporated on top of the poly-Si layer as a mask for the etching, and we use anisotropic reactive ion etching (RIE) to remove poly-Si in the regions that are not covered by Cr. Finally, we use wet etching to remove the Cr layer.

\section{Fabrication of nonlinear waveguide used for photon pair generation} \label{sec:fabNL} 

The orthogonally polarized photon pairs are generated from a type-II phase-matched periodically-poled lithium niobate (PPLN) waveguide. The PPLN grating period is around 9.4 $\mathrm{\mu}$m for a 785.3-nm pumped degenerate spontaneous parametric down-conversion (SPDC) process at $120^\circ \mathrm{C}$ (design value) according to the oeo quasi-phase matched (QPM) condition in Ti-diffused $\mathrm{LiNbO_3}$ single (fundamental)-mode waveguides. The fabrication process for the PPLN waveguides mainly involves Ti strips thermal in-diffusion followed by electric field poling in a $\mathrm{LiNbO_3}$ crystal~\cite{Myers:1995-2102:JOSB}.
The waveguides were fabricated using the titanium thermal diffusion (TTD) method~\cite{Chung:2015-30641:OE} in a 31-mm long, 10-mm wide, and 0.5-mm thick $z$-cut $\mathrm{LiNbO_3}$ crystal. Initially, an array of 7 $\mathrm{\mu}$m-wide, 90 nm-thick Ti strips was coated on the $-z$ surface of the crystal along the crystallographic $x$ axis using the standard lithographic and lift-off process. The titanium diffusion process was then performed in a high-temperature furnace at $1035^\circ \mathrm{C}$ with a constant oxygen flow for 12 hours. Before the crystal poling, optical polish on the $+z$ surface of the crystal was conducted to remove the domain-inverted layer on that surface formed during the high-temperature TTD process. Considering the possible fabrication errors, a QPM domain structure of multiple grating periods varying from 9.1 to 9.7 $\mathrm{\mu}$m, stepped by 0.1 $\mathrm{\mu}$m (along the crystallographic $y$ axis), was then implemented in the $\mathrm{LiNbO_3}$ waveguides using the standard electric field poling process. The device fabrication was then accomplished after the end faces of the crystal were optically polished.

\begin{figure*}[t]
	\centering
	\includegraphics[width=0.85\textwidth]{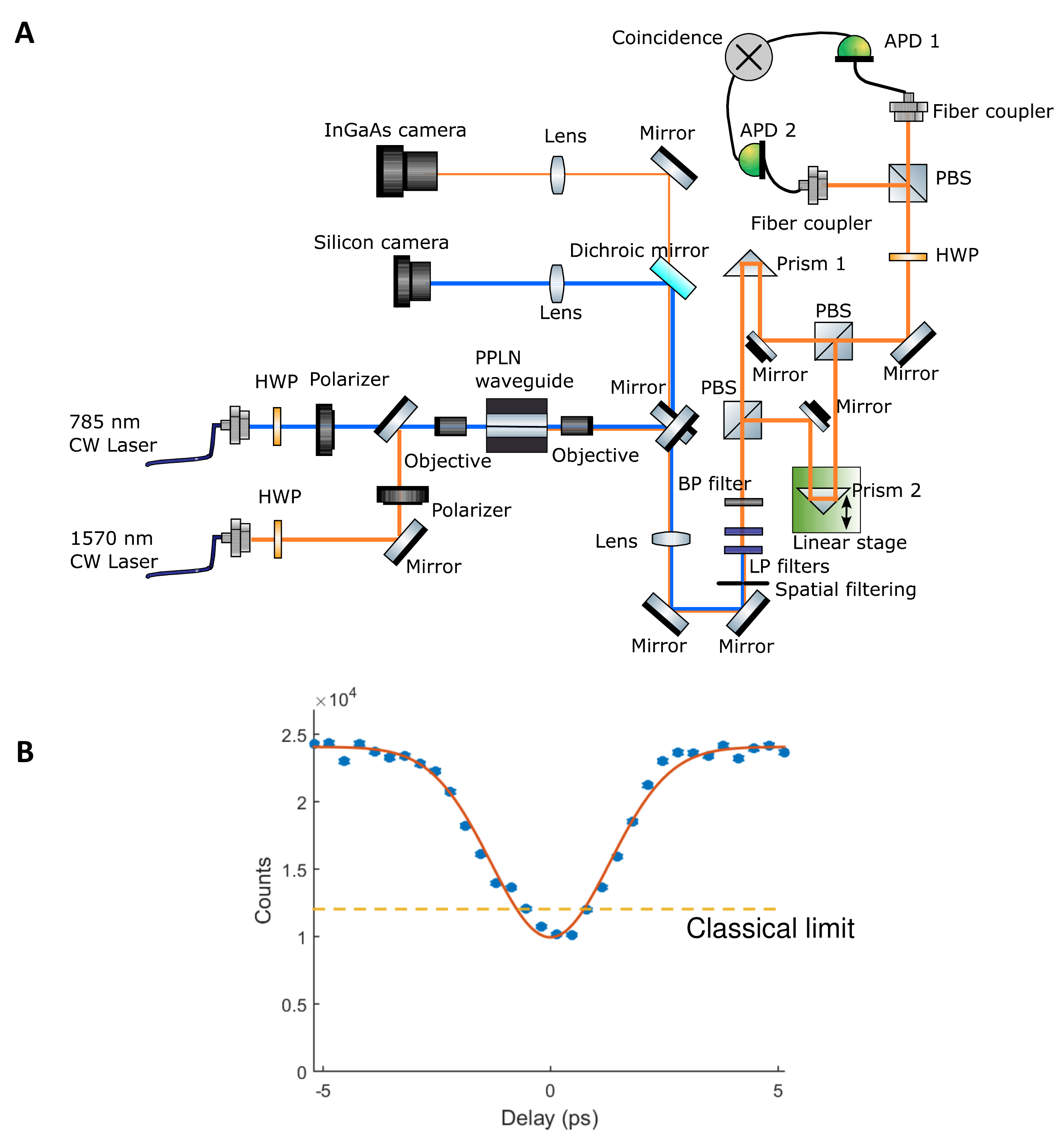}
	\caption{(A) Schematic setup for photon pair generation and interferometric characterizations, where HWP stands for half-wave-plate, PPLN is periodically poled lithium niobate, PBS is polarizing beam splitter, LP filter is long-pass filter (above 1100 nm can pass), BP filter is bandpass filter (1570 nm central wavelength with 12 nm FWHM) and QWP is quarter-wave-plate. (B) Experimental results of Hong-Ou-Mandel interference: the dots are experimental measurements and the solid curve represents a Gaussian fitting. The dashed line denotes half of the maximum counts given by the fitting, where a dip below this line exceeds the classical limit.
}\label{fig:hom}
\end{figure*}

\section{Influence of metasurface fabrication imperfections} \label{sec:imperfections} 

The most common type of fabrication imperfections may be associated with a scaled pattern in electron-beam lithography and under- or over-etching, where the effective dimensions of all nano-pillars would turn out a bit smaller or larger than designed. On the other hand, the overall metasurface geometry, such as the spacings between neighboring meta-atoms and the grating periodicity would be kept across the whole metasurface area. Then, the directions of the beam diffraction by the gratings will closely follow the design.

Therefore, the main effect of the fabrication inaccuracies would be in (i) a variation of elliptical polarization bases which are split between each pair of outputs and (ii) modification of diffraction efficiency, i.e. the fraction of photons directed to different outputs. Importantly, both the factors (i) and (ii) can be determined by performing an on-site characterization of the metasurface before it is used for quantum measurements. Specifically, we perform on-site measurements with classical light to determine the transfer matrix of the fabricated metasurface, as described below in \ref{sec:mschar} and \ref{sec:onSiteClassical}. Then the multi-photon transfer matrix is directly calculated as tensor products of the classical transfer matrices. By construction, the experimentally determined transfer matrix can incorporate all the features of fabrication imperfections, and therefore enables highly accurate reconstruction of the quantum states.
 
One minor influence of fabrication imperfection lies in a slightly higher amplification of errors in the reconstruction, as the experimentally achieved transfer matrix may be less optimal than the ideal theoretical design. 
As discussed in \ref{sec:of}, the condition number measures the extent that an error is amplified in the reconstruction. In the optimal case, the standard deviation of the error in the reconstruction is $\sqrt{3} \simeq 1.73$ times of the initial error in the measurement. In our experimentally characterized case this value is 2.08, which is only slightly higher and far from reaching the undesirable case of the condition number going to infinity.  This allows for robust reconstruction in the presence of shot noise [see Fig.~\ref{fig2}(C)].

Another influence of fabrication imperfections is the reduced efficiency. In the measurement of density matrices, non-perfect metasurface transmission and diffraction would effectively reduce the detector efficiency, without fundamentally affecting the measurement protocol -- the measurement only becomes more time-consuming. 
Specifically, we used single-photon detectors which have up to 25\% quantum efficiency, whereas our all-dielectric metasurface is tailored to be highly transparent (for an analysis of the high-transmission  metasurface designs see, e.g., Ref.~\cite{Kruk:2016-30801:APLP}).

\section{Metasurface characterization}~\label{sec:mschar}  

The fabricated metasurfaces are classically characterized using a tunable continues-wave laser. We measure the diffraction efficiency and characterize the projective bases across a broad bandwidth around the designed telecommunication wavelength. In Fig.~\ref{fig:classical}~(A) we show two example far-field images of the 6-output metasurface. The higher order diffractions along the vertical direction originate from the arrangement of metagratings, where the same grating is repeated 100 times before the next metagrating, in order to lower down the $k$-vector of diffractions in the vertical direction and utilize most of the light in the collection process (see \ref{sec:interleaving}). Thus practically these vertical diffractions can also be collected by the detector if one uses a cylindrical lens. If a camera (e.g. EMCCD) is used as the photon detector, then one can simply integrate multiple pixels in the vertical direction for efficient utilization of photons. The zero-order central spot in Fig.~\ref{fig:classical}~(A) is caused by fabrication errors and can be potentially eliminated by improving the fabrication. The example far-field images are compared to simulation. Due to the existing fabrication errors, 
the projective bases are not exactly the same as those designed theoretically.
However, this only has minor influence on the accuracy of the measurement result as the metasurface can be calibrated experimentally by classical light before performing quantum measurements (see \ref{sec:imperfections}). In Fig.~\ref{fig:classical}~(B) we show the projective bases of the 6-output metasurface across a broad bandwidth, illustrating that the metasurface can be utilized for measuring quantum light with large bandwidth -- here we characterized its projective bases for about 95~nm range in wavelength. In Figs.~\ref{fig:classical}(C) and (D) we show the diffraction efficiency of the metasurface for a 6-output sample and an 8-output sample, respectively. The 6-output sample is composed of three metagratings, with 8, 10, 14 meta-atoms in each periodicity, respectively (lattice parameter 800 nm). With the same lattice parameter, the 8-output metasurface is designed with four metagratings with 8, 10, 14, 20 meta-atoms in each periodicity, respectively. The diffraction efficiency here is characterized by the light power of the useful diffracted spots divided by the total power of all spots, collected by a microscope objective (NA 0.40) via $k$-space imaging. The characterization shows that the diffraction efficiency is higher than 85\% for both samples, indicating that our metasurfaces can enable very efficient manipulation and measurement of light, which is particularly important for quantum light that is typically weak. The minor fluctuations in the efficiency results in Fig.~~\ref{fig:classical}(D) can be explained by measurement errors caused by the dark image subtraction. In these classical measurements, a dark image (i.e. an image taken when the laser is turned off) was subtracted from the captured images. The dark image exhibits spatially-varying pattern, the overall brightness of which is affected by changes in the environmental light. Since the locations of diffracted spots vary with the wavelength, if the environmental light when the dark image was taken slightly deviates from that in the actual measurement, there can be a wavelength-dependent noise.

The power collected from each port is highly polarization dependent. Instead, we define a diffraction efficiency for a pair of ports diffracted from the same grating. Indeed, since each of the $M/2$ interleaved gratings occupies the same relative area of the metasurface, one grating should ideally diffract out $2/M$ fraction of the incident beam power to the corresponding pair of ports. We show such individual diffraction efficiencies in Fig.~\ref{fig:porteff}. We observe that light incident on the metasurface is almost equally distributed over different interleaved gratings across a wavelength region, staying close to the theoretical design efficiencies of $2/M$. The small fluctuating noise in these classically measured diffraction efficiencies with respect to wavelength can be explained by the varying environmental light and background subtraction, as the measurements are performed by scanning a tunable CW laser instead of launching broadband light source.

\begin{figure*}[t]
	\centering
	\includegraphics[width=1\textwidth]{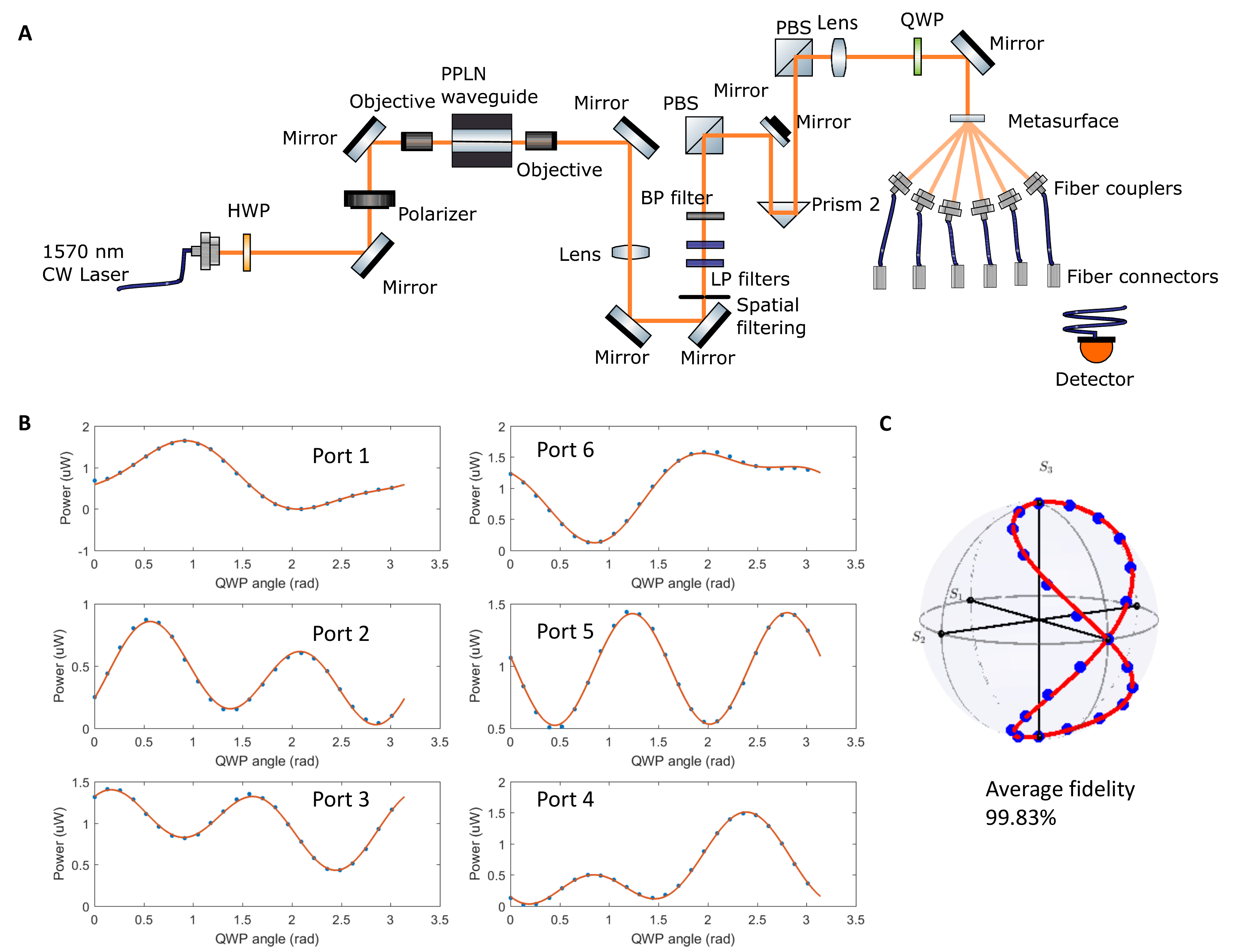}
	\caption{(A) Experimental setup for on-site classical measurements. (B) Powers measured from different ports after the metasurface (dots) compared to theoretical predications (curves) based on the on-site characterized transfer matrix. The error bars are not plotted as they are smaller than the marker size. (C) Reconstructed states (dots) plotted on the Poincare sphere and compared to the prepared states (curve).
}\label{fig:classicalSetup}
\end{figure*}

\section{SPDC waveguide characterization} \label{sec:SPDC} 

The characteristics of the fabricated waveguides were measured by using an external cavity laser ECL (Agilent, 8164B), whose wavelength is tunable from 1495 nm to 1600 nm. The waveguides support a single (fundamental) guiding mode for both horizontally (H) and vertically (V) polarized light in the laser wavelength tuning range. The measured waveguide propagation losses are approximately 0.3 and 0.6 dB/cm for V-polarized and H-polarized 1550 nm modes, respectively.
We then derived the photon-pair generation rate of our fabricated PPLN waveguide SPDC source via the study of the sum-frequency generation (SFG) process of the source based on the quantum-classical correspondence~\cite{Lenzini:2018-17143:LSA}. In the SFG measurement, we aligned a tunable laser to be 45-degree linearly polarized to provide two orthogonally (H- and V-) polarized modes in the input beam of the same wavelength, playing a role of the two pumps to the PPLN waveguide device. The measured QPM SFG temperature tuning rates are $-0.17$ to $-0.22$ nm/$^\circ \mathrm{C}$ (in the temperature range from 25$^\circ \mathrm{C}$ to 180$^\circ \mathrm{C}$) for SFG pump wavelengths ranging from 1540~nm to 1600~nm. This implies, for a 785 nm SPDC pump, the temperature for producing frequency-degenerate 1570.6~nm signal and idler photon pairs of 131.4$^\circ \mathrm{C}$ if a PPLN grating period of $\Lambda=9.4\ \mathrm{\mu m}$ is used. There is a shift in the phase-matching temperature between the measured and calculated results, which is a result of the deviation of the waveguide dispersion curve due to fabrication errors.
Furthermore, the measured results showed that the SFG bandwidth (FWHM) is ~0.43 nm and the SFG conversion efficiency under 319 $\mathrm{\mu}$W pump is about $4.33\times10^{-5}$, from which we estimate SPDC photon pair generation rate~\cite{Lenzini:2018-17143:LSA} under 1 mW 785-nm pump as ~$2.69\times10^7$ Hz when a 20-mm long, 9.4 $\mathrm{\mu}$m period PPLN waveguide is used.

\section{Photon pair source and its interferometric characterization}\label{sec:ppg} 

We use photon pairs generated via spontaneous parametric down conversion (SPDC) via a periodically poled lithium niobate (PPLN) waveguide as shown in Fig.~\ref{fig:hom}(A). The PPLN waveguide is heated up in a oven with closed-loop temperature control and is kept at $125^\circ \mathrm{C}$ in order to avoid photorefraction. The waveguide is pumped by a CW laser with a wavelength of $785.3$ nm, with about $10$ mW power launched into the waveguide by a microscope objective. The PPLN waveguide is designed for type-II phase matching, such that the horizontally (H) and vertically (V) polarized modes at around 1570.6 nm are phase matched. Therefore, we can obtain polarization entangled photon pairs with cross-polarized modes of H and V photons in a pair. The phase-matching is tunable by slightly changing the temperature, and we experimentally adjust it to degenerate frequencies for both photons in a pair. Since the H and V polarized modal profiles of photons in the waveguide are slightly different according to the transverse refractive index profile, we also perform a spatial filtering with a small iris [see Fig.~\ref{fig:hom}(A)] in order to select the well-overlapped and central region out of the transverse cross section of the photon beams, where both modes have approximately homogeneous intensity and polarization.

We apply the knowledge that 
type-II SPDC process is designed to be phase-matched and it generates 
a pair of photons which must be H and V cross-polarized, whereas the $HH$ or $VV$ photon pair generation processes are highly phase-mismatched and thus co-polarized photon pairs are practically never generated. Accordingly, we predict that only the central four elements in the two-photon density matrix (in $[HH, HV, VH, VV]$ basis) are non-zero. Moreover, with the symmetry of the density matrix due to the use of the indistinguishable detection scheme (see a related discussion on indistinguishable detection in the supplementary material section 1.2 of Ref.~\cite{Titchener:2018-19:NPJQI}), we theoretically predict the reduced density matrix of the photon-pair state to be
\begin{equation}\label{eq:dm0}
\mathbf{\rho}
=\frac{1}{2}\left[
\begin{array}{cccc}
0& 0& 0& 0 \\
0& 1& \eta& 0 \\
0& \eta& 1& 0 \\
0& 0& 0& 0 
\end{array}
\right],
\end{equation}
where $\eta$ is a real-valued number describing the spectral overlap of the photons in a pair (for details see supplementary material section 3 of Ref.~\cite{Titchener:2018-19:NPJQI}). 
The absolute value of $\eta$ is unity for photons with identical spectra, and accordingly identical temporal profiles.
On the other hand, in the HOM measurement the mismatched time delay between the photons results in the spectral phase tilt, and then $\eta$ approaches zero. Our state measurement experiment is performed when the time delay between the photons is best matched. At this point, it can be demonstrated that $\eta$ is exactly the depth of the dip in the HOM interference (i.e. the normalized HOM counts value is $1-\eta$). The experimentally measured HOM dip of our photon pair source is 58\% in depth with respect to the fully mismatched case, thus the theoretically predicted density matrix is
\begin{equation}\label{eq:dm}
\mathbf{\rho}
=\frac{1}{2}\left[
\begin{array}{cccc}
0& 0& 0& 0 \\
0& 1& 0.58& 0 \\
0& 0.58& 1& 0 \\
0& 0& 0& 0 
\end{array}
\right].
\end{equation}

We calculate the  concurrence~\cite{Wootters:1998-2245:PRL} of our prepared state, which is a measure between 0 and 1 of the degree of entanglement, as
\begin{equation}\label{eq:cc}
Cc=\max \{0, \lambda_1-\lambda_2-\lambda_3-\lambda_4\}=0.58,
\end{equation}
where $\lambda_is$ are the square roots of eigenvalues (in decreasing order) of $\mathbf{\rho\tilde{\rho}}$ with $\mathbf{\tilde{\rho}}=(\mathbf{\sigma_y}\otimes\mathbf{\sigma_y})\rho^\ast (\mathbf{\sigma_y}\otimes\mathbf{\sigma_y})$.
We also calculate the purity of state, 
\begin{equation}
P=\mathrm{Tr}(\mathbf{\rho}^2) [\mathrm{Tr}(\mathbf{\rho})]^{-2} =66.82\%.
\end{equation}
For such a four-dimensional state, $P$ can take the values ranging from 0.25 for fully mixed (incoherent) state to 1 for a pure (coherent) state. Thus the 66.82\% purity means that the state we use is a partially coherent state, which is also consistent with the intermediate degree of entanglement according to Eq.~\eqref{eq:cc}. Such a mixed state instead of an ideally pure one (i.e. $\eta$ does not approach unity at the HOM dip) is likely to be caused by the different spectra of the H and V photons and hence an imperfect spectral overlap even when the time delay between the H and V photons is matched. The joint spectra of the photon pairs are determined by the phase-matching condition of the SPDC waveguide and are affected by fabrication errors of the waveguide, in particular in the periodic poling. 

For reference, if the delay between the photons is not matched temporally (i.e. out of the dip in HOM), then a pair of photons cannot interfere but they are still cross-polarized (H and V). The reduced density matrix in this case is
\begin{equation}\label{eq:dmi}
\mathbf{\rho'}
=\frac{1}{2}\left[
\begin{array}{cccc}
0& 0& 0& 0 \\
0& 1& 0& 0 \\
0& 0& 1& 0 \\
0& 0& 0& 0 
\end{array}
\right].
\end{equation}

\begin{figure*}[t]
	\centering
	\includegraphics[width=1\textwidth]{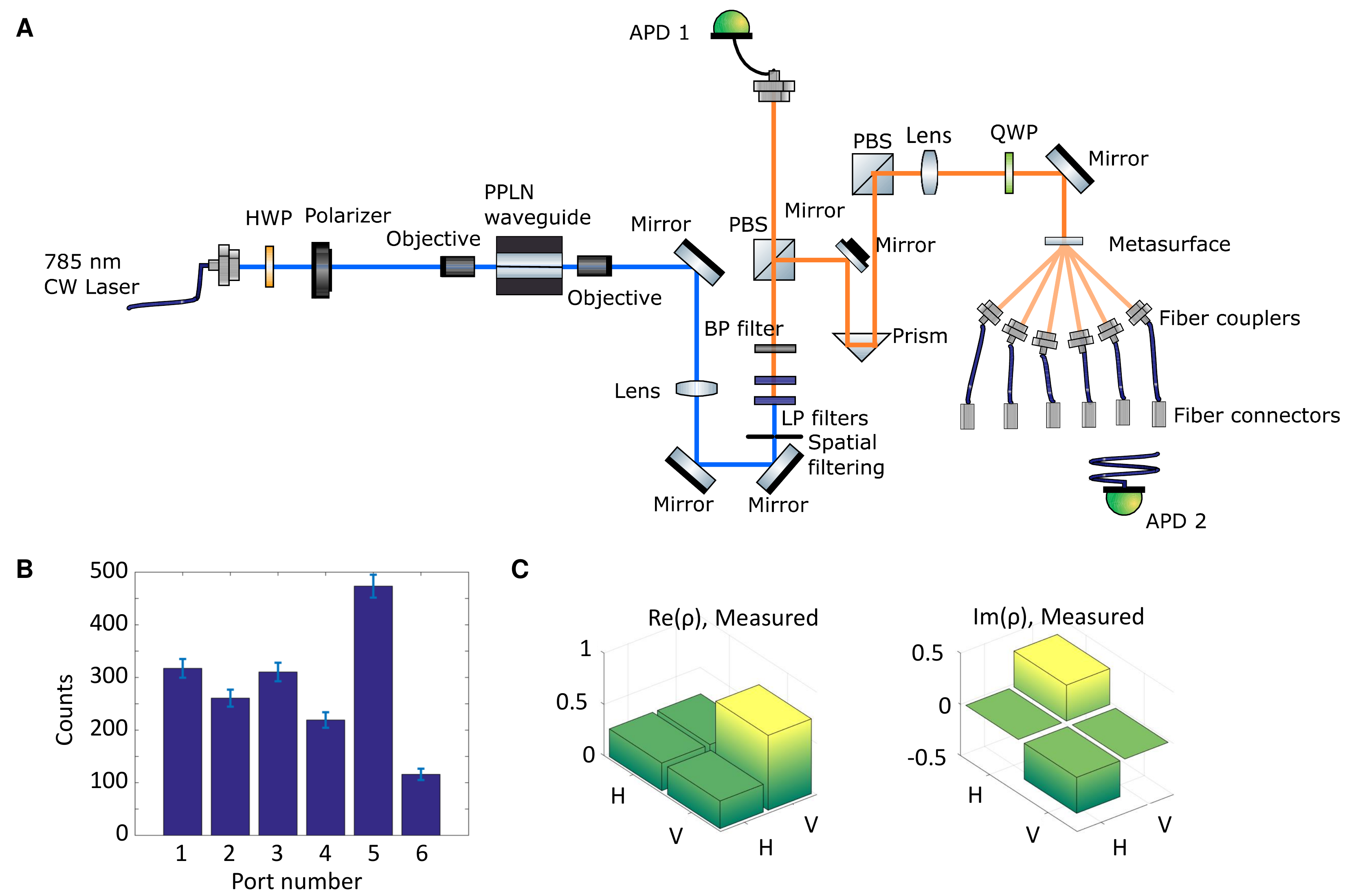}
	\caption{(A) Setup for heralded single-photon measurements, where APD stands for avalanche photodiode that is used as our single-photon detector. The two APDs are connected to time-to-digital conversion electronics to record the correlation of events from the two detectors (not shown in the figure). (B) A representative readout from the single-photon measurement, where the error bars are based on the shot noise in photon detection. (C) the reconstructed single-photon density matrix $\mathbf{\rho}$ using the measurement data in (B). 
}\label{fig:sp}
\end{figure*}

\section{Experimental scheme for on-site classical measurements} \label{sec:onSiteClassical} 

As discussed in \ref{sec:mschar} above, we initially performed a classical characterization of the metasurface in an imaging-based setup, and its transfer matrix has been obtained across a broad band. Our quantum measurements are based on a different setup due to the implementation of photon pair source and fiber-coupled single-photon detection interface. The actual transfer matrix does not only depend on the metasurface, but is also slightly influenced by the spatial mode of light and coupling conditions into the fiber interface that connects to the detectors. Therefore we also perform an on-site classical measurement to characterize the actual transfer matrix that is exactly the same for quantum measurements and to verify that the system works properly. The setup for such an on-site measurement is shown in Fig.~\ref{fig:classicalSetup}(A). A CW laser at 1570.6 nm wavelength is firstly linearly polarized by a Glan-Taylor prism (polarizer) and then launched into the waveguide which we use for photon pair generation. Starting from the SPDC waveguide, the laser beam goes through practically the same optical elements as the V-polarized photons in the pair of the quantum light,
to ensure a similar spatial mode. Then we transform the V polarized component of the laser beam with a quarter-wave-plate (QWP) to prepare many different states and slightly reduce the waist of the laser beam by a lens. The spot of the laser beam illuminated on the metasurface has a diameter of approximately 1~mm, which is smaller than the size of the metasurface ($\mathrm{2\ mm \times 2\ mm}$).
We then use an IR detector to read out the power from each fiber port. By doing so, the transfer matrix is fully characterized. Then, we prepare another set of polarization states by the same QWP, and reconstruct the input polarization states using the transfer matrix. The experimental results of powers measured from each port are shown in Fig.~\ref{fig:classicalSetup} (B) with the dots for experimental data and the curves for theoretical predictions by the characterized transfer matrix. In Fig.~\ref{fig:classicalSetup} (C) we show the reconstructed states on a Poincar\'e sphere. The fidelity is very high, reaching an average value of 99.83\%. Such a fidelity is higher than for single-photon measurements reported in Fig.~\ref{fig2}, since classical measurements are not limited by the shot noise.  

The characterized transfer matrix of the metasurface is
\begin{equation}
\mathbf{T}
= \xi \left[
\begin{array}{ccc}
 1.000  & -0.3227 - 0.7070i \\
   1.2022 + 0.2874i & 0.6484 \\
  0.1781 + 0.1282i &  0.7935 \\
   -0.2692 - 0.8502i &  0.2683 \\
   -0.6830 + 0.0063i &  0.8625 \\
  0.1971 - 0.5392i &  1.1189 
\end{array}
\right],
\end{equation}
where $\xi$ is a scaling factor and each row is essentially a projective basis of one of the 6 spots from the metasurface for classical or single-photon measurements. The two columns denote H and V polarizations, respectively. Note that each row can be gauge transformed up to a phase factor. The $N$-photon transfer matrix is basically the Kronecker product of $\mathbf{T}$ as $\mathbf{T}^{\otimes N}=\mathbf{T}\otimes\cdots \otimes \mathbf{T}$.

\begin{figure*}[t]
	\centering
	\includegraphics[width=\textwidth]{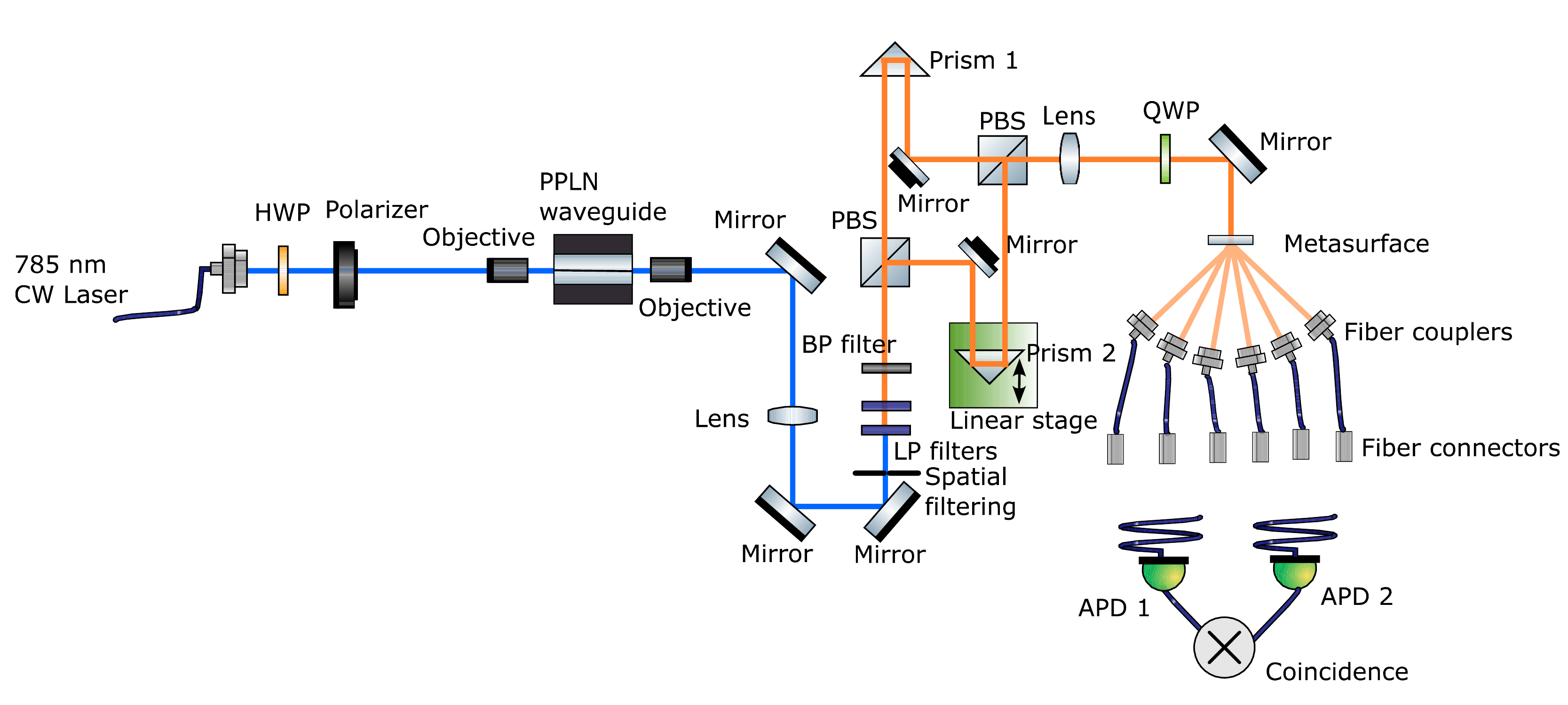}
	\caption{Schematic setup for heralded single-photon measurements. We use a motorized linear stage with a 25 mm range for varying the time delay between H and V photons. 
}\label{fig:tp}
\end{figure*}

\begin{figure*}
	\centering
	\includegraphics[width=1\textwidth]{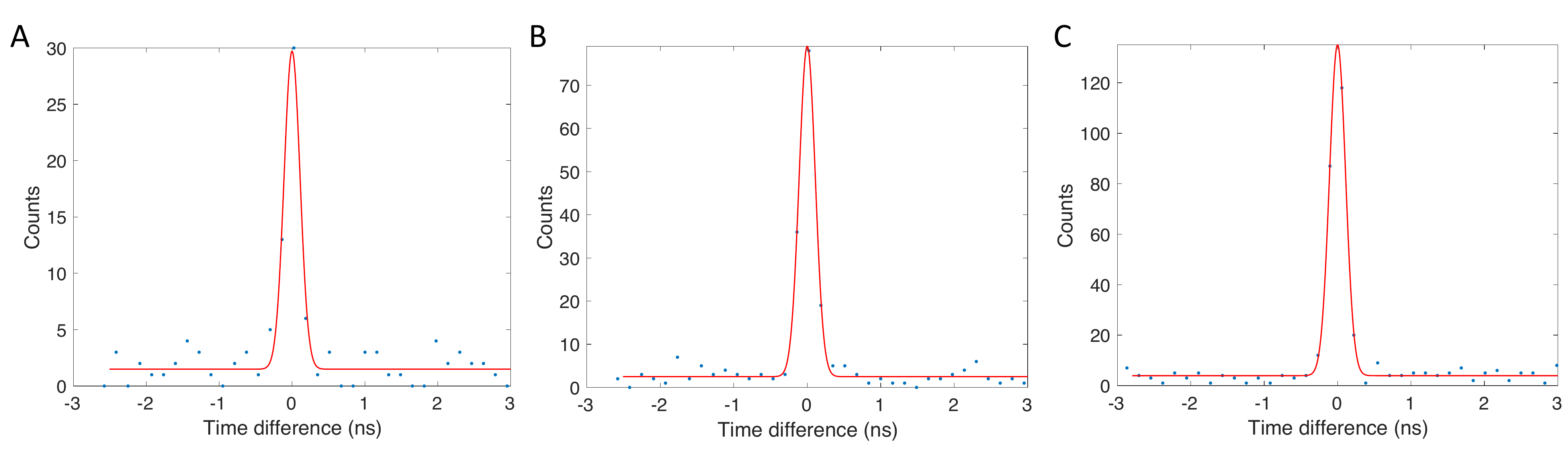}
	\caption{Representative correlation histograms (dots) of two-photon experimental measurements corresponding to  Fig.~\ref{fig3}(F) for the port combinations (A) 1-3 (B) 1-4 and (C) 2-6. The red curve denotes a Gaussian fitting with an offset in counts for the removal of background.}\label{fig:hist}
\end{figure*}

\section{Experimental scheme for heralded single-photon experiment} \label{sec:expHeralded} 

For the single-photon experiment, we use the cross-polarized photon pairs generated with the waveguide as described above in \ref{sec:ppg} to implement a heralded single-photon source. The schematic of the setup is shown in Fig.~\ref{fig:sp}~(A). Different from the setup shown in Fig.~\ref{fig:hom}(A), after the pump-filtered H and V photons are separated by a polarizing beam splitter (PBS), the H photons are sent to a single-photon detector (APD 1) as the master arm in the photon heralding. The V photons go through a QWP and then are sent to the metasurface, and photons at the six diffracted spots are collected by the fiber couplers. Each fiber can be connected to another single-photon detector (APD 2), which acts as the slave detector. To efficiently collect the photons in the quantum measurements, we use cylindrical lenses in between the metasurface and the fiber couplers to focus the vertically diffracted photons (not shown in the figure). Since the SPDC process in our experiment has very low probably to generate photon states with higher than 2 photons, if both APD 1 and APD 2 click we can be sure that this event on APD 2 is a single-photon event, which is the so-called heralded single-photon. In Fig.~\ref{fig:sp}~(B), we show a representative readout of such correlation counts between APD 1 and APD 2, where APD 2 is switched between different output ports (1--6). With such counts, we are able to reconstruct both the real and imaginary parts of the single-photon density matrix $\mathbf{\rho}$, as shown in Fig.~\ref{fig:sp}(C). The reconstructed state is consistent with the theoretical predictions.

\section{Experimental scheme for two-photon experiment} \label{sec:expTwoPhoton}  

A more detailed setup of two-photon experiment is presented in Fig.~\ref{fig:tp}. The setup is analogous to the one described above in \ref{sec:ppg} for the SPDC generation and characterization of photon pairs via HOM interference. After the photon pairs go through a delay-line that can vary the time delay between the two photons, they are transformed by a QWP and then directly launched onto the metasurface. The 6 diffracted beams from the metasurface are collected by fiber couplers into multi-mode fibers. In the measurement, we switch the fiber connectors with two single-photon detectors and map out all combinations of two out of the six output ports (15 measurements).

Here we present the procedure for the prediction of the quantum interference behavior of our two photon state for different combinations of two ports. Representatively, if we measure photon interference between port 1 and port 6, the $2\times2$ transfer matrix is
\begin{equation}
\begin{split}
\mathbf{T_{16}}
&= \xi \left[ \mathbf{u_1}, \mathbf{u_6}\right]^\dagger \\
&=\xi \left[
\begin{array}{ccc}
   1.000 &  -0.3227 - 0.7070i \\
  0.1971 - 0.5392i & 1.1189 
 \end{array}
\right],
\end{split}
\end{equation}
where $\xi$ is a scaling factor. For the following analysis we take $\xi=1$ and hence the theoretically predicated measurement expectation values are in arbitrary units. 
For our two-photon density matrix given in Eq.~\eqref{eq:dm}, we can calculate the expectation value of two-fold coincidence in the measurement by
\begin{equation}
C_{16}= 2 \left[\mathbf{u_1} \otimes \mathbf{u_6}\right]^\dagger \mathbf{\rho} \left[\mathbf{u_1} \otimes \mathbf{u_6}\right]=0.8737,
\end{equation}
which corresponds to the correlation counts when the time delay of H and V photons is matched. For the case that the two photons are fully mismatched in time, following the density matrix given in Eq.~\eqref{eq:dmi}, we have
\begin{equation}
C'_{16}= 2 \left[\mathbf{u_1} \otimes \mathbf{u_6}\right]^\dagger \mathbf{\rho'} \left[\mathbf{u_1} \otimes \mathbf{u_6}\right] =1.4511.
\end{equation}
Since $C_{16}<C'_{16}$, theoretically we predict that in the quantum interference using ports 1 and 6, one obtains a dip with a relative depth of $(C'_{16}-C_{16})/C'_{16}= 39.79\%$. Such a dip is plotted with a solid curve in Fig.~\ref{fig3}(B) of the main text as a theoretical prediction.

Similarly, we can run the calculations for ports 1 and 5, with their transfer matrix being
\begin{equation}
\begin{split}
\mathbf{T_{16}}
&= \xi \left[ \mathbf{u_1}, \mathbf{u_5}\right]^\dagger \\
&=\xi \left[
\begin{array}{ccc}
   1.000 &  -0.3227 - 0.7070i \\
  -0.6830 + 0.0063i &  0.8625
 \end{array}
\right].
\end{split}
\end{equation}
Hence the expectation of two-fold correlation when the time delay of H and V photons is matched can be given by
\begin{equation}
C_{15}= 2 \left[\mathbf{u_1} \otimes \mathbf{u_5}\right]^\dagger \mathbf{\rho} \left[\mathbf{u_1} \otimes \mathbf{u_5}\right]=1.2505,
\end{equation}
For temporally mismatched photons, 
\begin{equation}
C'_{15}= 2 \left[\mathbf{u_1} \otimes \mathbf{u_5}\right]^\dagger \mathbf{\rho'} \left[\mathbf{u_1} \otimes \mathbf{u_5}\right] =1.0256.
\end{equation}
Since $C_{15}>C'_{15}$, theoretically we predict that in the quantum interference using ports 1 and 5, one obtains a peak with a relative hight of $|C'_{15}-C_{15}|/C'_{15}= 21.93\%$. In Fig.~\ref{fig3}(C) of the main text, we plot such a peak with a solid curve as our theoretical prediction and confirm the consistency with the experimental measurements. 

For the quantum correlation measurements, including both the heralded single photon and two-photon measurements, we obtain the histograms, and the correlation counts are extracted by processing these histograms. Note that there is essentially a background subtraction as we use a Gaussian fitting to the correlation histogram. The background is no more than 10\% even for very low counts with respect to the signal, mainly originating from dark counts and accidental counts. In Fig.~\ref{fig:hist} we show three representative histograms from the two-photon measurements shown in Fig.~\ref{fig3}(F). 

\begin{figure*}
	\centering
	\includegraphics[width=1\textwidth]{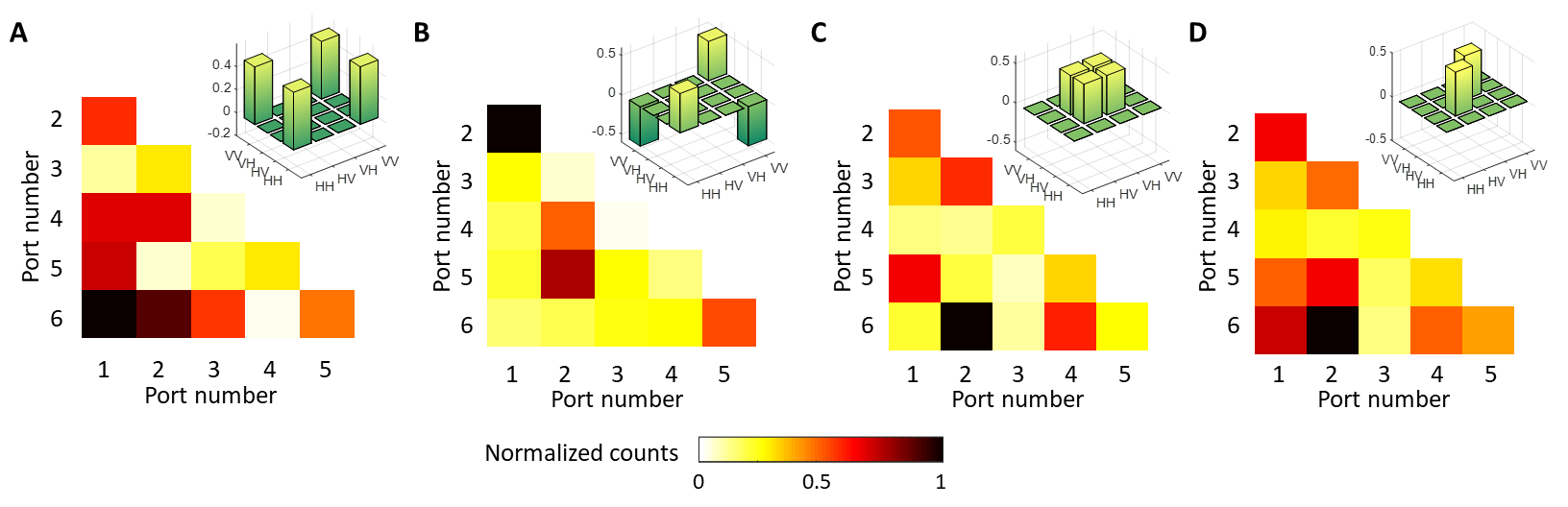}
	\caption{Theoretically predicted two-fold correlation counts based on experimentally characterized transfer matrix of the metasurface, shown for representative two-photon polarization states: (A) bunched state, (B) bunched state with $\pi$ phase difference, (C) anti-bunched state, (D) incoherent anti-bunched state. The real-parts of the input density matrices are shown on the top right in the respective sub-figures.}\label{fig:corr}
\end{figure*}

Additionally, for a more intuitive understanding of the two-photon state measurement, in Fig.~\ref{fig:corr} we plot the correlations which would be produced by different pure and mixed states, calculated based on the experimentally measured transfer matrix of the metasurface. Since the transfer matrix corresponds to projections onto elliptical polarization bases, one can see that even pure states defined in linear polarization bases would not have vanishing counts at any ports.


\begin{thebibliography}{36}%
\makeatletter
\providecommand \@ifxundefined [1]{%
 \@ifx{#1\undefined}
}%
\providecommand \@ifnum [1]{%
 \ifnum #1\expandafter \@firstoftwo
 \else \expandafter \@secondoftwo
 \fi
}%
\providecommand \@ifx [1]{%
 \ifx #1\expandafter \@firstoftwo
 \else \expandafter \@secondoftwo
 \fi
}%
\providecommand \natexlab [1]{#1}%
\providecommand \enquote  [1]{``#1''}%
\providecommand \bibnamefont  [1]{#1}%
\providecommand \bibfnamefont [1]{#1}%
\providecommand \citenamefont [1]{#1}%
\providecommand \href@noop [0]{\@secondoftwo}%
\providecommand \href [0]{\begingroup \@sanitize@url \@href}%
\providecommand \@href[1]{\@@startlink{#1}\@@href}%
\providecommand \@@href[1]{\endgroup#1\@@endlink}%
\providecommand \@sanitize@url [0]{\catcode `\\12\catcode `\$12\catcode
  `\&12\catcode `\#12\catcode `\^12\catcode `\_12\catcode `\%12\relax}%
\providecommand \@@startlink[1]{}%
\providecommand \@@endlink[0]{}%
\providecommand \url  [0]{\begingroup\@sanitize@url \@url }%
\providecommand \@url [1]{\endgroup\@href {#1}{\urlprefix }}%
\providecommand \urlprefix  [0]{URL }%
\providecommand \Eprint [0]{\href }%
\providecommand \doibase [0]{http://dx.doi.org/}%
\providecommand \selectlanguage [0]{\@gobble}%
\providecommand \bibinfo  [0]{\@secondoftwo}%
\providecommand \bibfield  [0]{\@secondoftwo}%
\providecommand \translation [1]{[#1]}%
\providecommand \BibitemOpen [0]{}%
\providecommand \bibitemStop [0]{}%
\providecommand \bibitemNoStop [0]{.\EOS\space}%
\providecommand \EOS [0]{\spacefactor3000\relax}%
\providecommand \BibitemShut  [1]{\csname bibitem#1\endcsname}%
\let\auto@bib@innerbib\@empty
\bibitem [{\citenamefont {Yu}\ and\ \citenamefont
  {Capasso}(2014)}]{Yu:2014-139:NMAT}%
  \BibitemOpen
  \bibfield  {author} {\bibinfo {author} {\bibfnamefont {N.~F.}\ \bibnamefont
  {Yu}}\ and\ \bibinfo {author} {\bibfnamefont {F.}~\bibnamefont {Capasso}},\
  }\bibfield  {title} {{\selectlanguage {English}\enquote {\bibinfo {title}
  {Flat optics with designer metasurfaces},}\ }}\href {\doibase
  10.1038/nmat3839} {\bibfield  {journal} {\bibinfo  {journal} {Nat. Mater.}\
  }\textbf {\bibinfo {volume} {13}},\ \bibinfo {pages} {139--150} (\bibinfo
  {year} {2014})}\BibitemShut {NoStop}%
\bibitem [{\citenamefont {Kuznetsov}\ \emph {et~al.}(2016)\citenamefont
  {Kuznetsov}, \citenamefont {Miroshnichenko}, \citenamefont {Brongersma},
  \citenamefont {Kivshar},\ and\ \citenamefont
  {Luk'yanchuk}}]{Kuznetsov:2016-846:SCI}%
  \BibitemOpen
  \bibfield  {author} {\bibinfo {author} {\bibfnamefont {A.~I.}\ \bibnamefont
  {Kuznetsov}}, \bibinfo {author} {\bibfnamefont {A.~E.}\ \bibnamefont
  {Miroshnichenko}}, \bibinfo {author} {\bibfnamefont {M.~L.}\ \bibnamefont
  {Brongersma}}, \bibinfo {author} {\bibfnamefont {{Yu}.~S.}\ \bibnamefont
  {Kivshar}}, \ and\ \bibinfo {author} {\bibfnamefont {B.}~\bibnamefont
  {Luk'yanchuk}},\ }\bibfield  {title} {{\selectlanguage {English}\enquote
  {\bibinfo {title} {Optically resonant dielectric nanostructures},}\ }}\href
  {\doibase 10.1126/science.aag2472} {\bibfield  {journal} {\bibinfo  {journal}
  {Science}\ }\textbf {\bibinfo {volume} {354}},\ \bibinfo {pages} {846--854}
  (\bibinfo {year} {2016})}\BibitemShut {NoStop}%
\bibitem [{\citenamefont {Decker}\ \emph {et~al.}(2015)\citenamefont {Decker},
  \citenamefont {Staude}, \citenamefont {Falkner}, \citenamefont {Dominguez},
  \citenamefont {Neshev}, \citenamefont {Brener}, \citenamefont {Pertsch},\
  and\ \citenamefont {Kivshar}}]{Decker:2015-813:ADOM}%
  \BibitemOpen
  \bibfield  {author} {\bibinfo {author} {\bibfnamefont {M.}~\bibnamefont
  {Decker}}, \bibinfo {author} {\bibfnamefont {I.}~\bibnamefont {Staude}},
  \bibinfo {author} {\bibfnamefont {M.}~\bibnamefont {Falkner}}, \bibinfo
  {author} {\bibfnamefont {J.}~\bibnamefont {Dominguez}}, \bibinfo {author}
  {\bibfnamefont {D.~N.}\ \bibnamefont {Neshev}}, \bibinfo {author}
  {\bibfnamefont {I.}~\bibnamefont {Brener}}, \bibinfo {author} {\bibfnamefont
  {T.}~\bibnamefont {Pertsch}}, \ and\ \bibinfo {author} {\bibfnamefont
  {{Yu}.~S.}\ \bibnamefont {Kivshar}},\ }\bibfield  {title} {{\selectlanguage
  {English}\enquote {\bibinfo {title} {High-efficiency dielectric huygens'
  surfaces},}\ }}\href {\doibase 10.1002/adom.201400584} {\bibfield  {journal}
  {\bibinfo  {journal} {Adv. Opt. Mater.}\ }\textbf {\bibinfo {volume} {3}},\
  \bibinfo {pages} {813--820} (\bibinfo {year} {2015})}\BibitemShut {NoStop}%
\bibitem [{\citenamefont {Arbabi}\ \emph {et~al.}(2015)\citenamefont {Arbabi},
  \citenamefont {Horie}, \citenamefont {Bagheri},\ and\ \citenamefont
  {Faraon}}]{Arbabi:2015-937:NNANO}%
  \BibitemOpen
  \bibfield  {author} {\bibinfo {author} {\bibfnamefont {A.}~\bibnamefont
  {Arbabi}}, \bibinfo {author} {\bibfnamefont {Y.}~\bibnamefont {Horie}},
  \bibinfo {author} {\bibfnamefont {M.}~\bibnamefont {Bagheri}}, \ and\
  \bibinfo {author} {\bibfnamefont {A.}~\bibnamefont {Faraon}},\ }\bibfield
  {title} {{\selectlanguage {English}\enquote {\bibinfo {title} {Dielectric
  metasurfaces for complete control of phase and polarization with
  subwavelength spatial resolution and high transmission},}\ }}\href {\doibase
  10.1038/NNANO.2015.186} {\bibfield  {journal} {\bibinfo  {journal} {Nat.
  Nanotechnol.}\ }\textbf {\bibinfo {volume} {10}},\ \bibinfo {pages}
  {937--U190} (\bibinfo {year} {2015})}\BibitemShut {NoStop}%
\bibitem [{\citenamefont {Kruk}\ \emph {et~al.}(2016)\citenamefont {Kruk},
  \citenamefont {Hopkins}, \citenamefont {Kravchenko}, \citenamefont
  {Miroshnichenko}, \citenamefont {Neshev},\ and\ \citenamefont
  {Kivshar}}]{Kruk:2016-30801:APLP}%
  \BibitemOpen
  \bibfield  {author} {\bibinfo {author} {\bibfnamefont {S.}~\bibnamefont
  {Kruk}}, \bibinfo {author} {\bibfnamefont {B.}~\bibnamefont {Hopkins}},
  \bibinfo {author} {\bibfnamefont {I.~I.}\ \bibnamefont {Kravchenko}},
  \bibinfo {author} {\bibfnamefont {A.}~\bibnamefont {Miroshnichenko}},
  \bibinfo {author} {\bibfnamefont {D.~N.}\ \bibnamefont {Neshev}}, \ and\
  \bibinfo {author} {\bibfnamefont {{Yu}.~S.}\ \bibnamefont {Kivshar}},\
  }\bibfield  {title} {{\selectlanguage {English}\enquote {\bibinfo {title}
  {Invited article: Broadband highly efficient dielectric metadevices for
  polarization control},}\ }}\href {\doibase 10.1063/1.4949007} {\bibfield
  {journal} {\bibinfo  {journal} {APL Photonics}\ }\textbf {\bibinfo {volume}
  {1}},\ \bibinfo {pages} {030801--9} (\bibinfo {year} {2016})}\BibitemShut
  {NoStop}%
\bibitem [{\citenamefont {Genevet}\ \emph {et~al.}(2017)\citenamefont
  {Genevet}, \citenamefont {Capasso}, \citenamefont {Aieta}, \citenamefont
  {Khorasaninejad},\ and\ \citenamefont {Devlin}}]{Genevet:2017-139:OPT}%
  \BibitemOpen
  \bibfield  {author} {\bibinfo {author} {\bibfnamefont {P.}~\bibnamefont
  {Genevet}}, \bibinfo {author} {\bibfnamefont {F.}~\bibnamefont {Capasso}},
  \bibinfo {author} {\bibfnamefont {F.}~\bibnamefont {Aieta}}, \bibinfo
  {author} {\bibfnamefont {M.}~\bibnamefont {Khorasaninejad}}, \ and\ \bibinfo
  {author} {\bibfnamefont {R.}~\bibnamefont {Devlin}},\ }\bibfield  {title}
  {\enquote {\bibinfo {title} {Recent advances in planar optics: from plasmonic
  to dielectric metasurfaces},}\ }\href {\doibase 10.1364/OPTICA.4.000139}
  {\bibfield  {journal} {\bibinfo  {journal} {Optica}\ }\textbf {\bibinfo
  {volume} {4}},\ \bibinfo {pages} {139--152} (\bibinfo {year}
  {2017})}\BibitemShut {NoStop}%
\bibitem [{\citenamefont {Mueller}\ \emph {et~al.}(2017)\citenamefont
  {Mueller}, \citenamefont {Rubin}, \citenamefont {Devlin}, \citenamefont
  {Groever},\ and\ \citenamefont {Capasso}}]{Mueller:2017-113901:PRL}%
  \BibitemOpen
  \bibfield  {author} {\bibinfo {author} {\bibfnamefont {J.~P.~B.}\
  \bibnamefont {Mueller}}, \bibinfo {author} {\bibfnamefont {N.~A.}\
  \bibnamefont {Rubin}}, \bibinfo {author} {\bibfnamefont {R.~C.}\ \bibnamefont
  {Devlin}}, \bibinfo {author} {\bibfnamefont {B.}~\bibnamefont {Groever}}, \
  and\ \bibinfo {author} {\bibfnamefont {F.}~\bibnamefont {Capasso}},\
  }\bibfield  {title} {{\selectlanguage {English}\enquote {\bibinfo {title}
  {Metasurface polarization optics: Independent phase control of arbitrary
  orthogonal states of polarization},}\ }}\href {\doibase
  10.1103/PhysRevLett.118.113901} {\bibfield  {journal} {\bibinfo  {journal}
  {Phys. Rev. Lett.}\ }\textbf {\bibinfo {volume} {118}},\ \bibinfo {pages}
  {113901--5} (\bibinfo {year} {2017})}\BibitemShut {NoStop}%
\bibitem [{\citenamefont {Bomzon}\ \emph {et~al.}(2001)\citenamefont {Bomzon},
  \citenamefont {Kleiner},\ and\ \citenamefont {Hasman}}]{Bomzon:2001-1424:OL}%
  \BibitemOpen
  \bibfield  {author} {\bibinfo {author} {\bibfnamefont {Z.}~\bibnamefont
  {Bomzon}}, \bibinfo {author} {\bibfnamefont {V.}~\bibnamefont {Kleiner}}, \
  and\ \bibinfo {author} {\bibfnamefont {E.}~\bibnamefont {Hasman}},\
  }\bibfield  {title} {{\selectlanguage {English}\enquote {\bibinfo {title}
  {Pancharatnam-berry phase in space-variant polarization-state manipulations
  with subwavelength gratings},}\ }}\href {\doibase 10.1364/OL.26.001424}
  {\bibfield  {journal} {\bibinfo  {journal} {Opt. Lett.}\ }\textbf {\bibinfo
  {volume} {26}},\ \bibinfo {pages} {1424--1426} (\bibinfo {year}
  {2001})}\BibitemShut {NoStop}%
\bibitem [{\citenamefont {Pors}\ \emph {et~al.}(2015)\citenamefont {Pors},
  \citenamefont {Nielsen},\ and\ \citenamefont
  {Bozhevolnyi}}]{Pors:2015-716:OPT}%
  \BibitemOpen
  \bibfield  {author} {\bibinfo {author} {\bibfnamefont {A.}~\bibnamefont
  {Pors}}, \bibinfo {author} {\bibfnamefont {M.~G.}\ \bibnamefont {Nielsen}}, \
  and\ \bibinfo {author} {\bibfnamefont {S.~I.}\ \bibnamefont {Bozhevolnyi}},\
  }\bibfield  {title} {{\selectlanguage {English}\enquote {\bibinfo {title}
  {Plasmonic metagratings for simultaneous determination of {S}tokes
  parameters},}\ }}\href {\doibase 10.1364/OPTICA.2.000716} {\bibfield
  {journal} {\bibinfo  {journal} {Optica}\ }\textbf {\bibinfo {volume} {2}},\
  \bibinfo {pages} {716--723} (\bibinfo {year} {2015})}\BibitemShut {NoStop}%
\bibitem [{\citenamefont {Mueller}\ \emph {et~al.}(2016)\citenamefont
  {Mueller}, \citenamefont {Leosson},\ and\ \citenamefont
  {Capasso}}]{Mueller:2016-42:OPT}%
  \BibitemOpen
  \bibfield  {author} {\bibinfo {author} {\bibfnamefont {J.~P.~B.}\
  \bibnamefont {Mueller}}, \bibinfo {author} {\bibfnamefont {K.}~\bibnamefont
  {Leosson}}, \ and\ \bibinfo {author} {\bibfnamefont {F.}~\bibnamefont
  {Capasso}},\ }\bibfield  {title} {{\selectlanguage {English}\enquote
  {\bibinfo {title} {Ultracompact metasurface in-line polarimeter},}\ }}\href
  {\doibase 10.1364/OPTICA.3.000042} {\bibfield  {journal} {\bibinfo  {journal}
  {Optica}\ }\textbf {\bibinfo {volume} {3}},\ \bibinfo {pages} {42--47}
  (\bibinfo {year} {2016})}\BibitemShut {NoStop}%
\bibitem [{\citenamefont {Maguid}\ \emph {et~al.}(2016)\citenamefont {Maguid},
  \citenamefont {Yulevich}, \citenamefont {Veksler}, \citenamefont {Kleiner},
  \citenamefont {Brongersma},\ and\ \citenamefont
  {Hasman}}]{Maguid:2016-1202:SCI}%
  \BibitemOpen
  \bibfield  {author} {\bibinfo {author} {\bibfnamefont {E.}~\bibnamefont
  {Maguid}}, \bibinfo {author} {\bibfnamefont {I.}~\bibnamefont {Yulevich}},
  \bibinfo {author} {\bibfnamefont {D.}~\bibnamefont {Veksler}}, \bibinfo
  {author} {\bibfnamefont {V.}~\bibnamefont {Kleiner}}, \bibinfo {author}
  {\bibfnamefont {M.~L.}\ \bibnamefont {Brongersma}}, \ and\ \bibinfo {author}
  {\bibfnamefont {E.}~\bibnamefont {Hasman}},\ }\bibfield  {title}
  {{\selectlanguage {English}\enquote {\bibinfo {title} {Photonic
  spin-controlled multifunctional shared-aperture antenna array},}\ }}\href
  {\doibase 10.1126/science.aaf3417} {\bibfield  {journal} {\bibinfo  {journal}
  {Science}\ }\textbf {\bibinfo {volume} {352}},\ \bibinfo {pages} {1202--1206}
  (\bibinfo {year} {2016})}\BibitemShut {NoStop}%
\bibitem [{\citenamefont {Ding}\ \emph {et~al.}(2017)\citenamefont {Ding},
  \citenamefont {Pors}, \citenamefont {Chen}, \citenamefont {Zenin},\ and\
  \citenamefont {Bozhevolnyi}}]{Ding:2017-943:ACSP}%
  \BibitemOpen
  \bibfield  {author} {\bibinfo {author} {\bibfnamefont {F.}~\bibnamefont
  {Ding}}, \bibinfo {author} {\bibfnamefont {A.}~\bibnamefont {Pors}}, \bibinfo
  {author} {\bibfnamefont {Y.~T.}\ \bibnamefont {Chen}}, \bibinfo {author}
  {\bibfnamefont {V.~A.}\ \bibnamefont {Zenin}}, \ and\ \bibinfo {author}
  {\bibfnamefont {S.~I.}\ \bibnamefont {Bozhevolnyi}},\ }\bibfield  {title}
  {{\selectlanguage {English}\enquote {\bibinfo {title} {Beam-size-invariant
  spectropolarimeters using gap-plasmon metasurfaces},}\ }}\href {\doibase
  10.1021/acsphotonics.6b01046} {\bibfield  {journal} {\bibinfo  {journal} {ACS
  Photonics}\ }\textbf {\bibinfo {volume} {4}},\ \bibinfo {pages} {943--949}
  (\bibinfo {year} {2017})}\BibitemShut {NoStop}%
\bibitem [{\citenamefont {Devlin}\ \emph {et~al.}(2017)\citenamefont {Devlin},
  \citenamefont {Ambrosio}, \citenamefont {Rubin}, \citenamefont {Mueller},\
  and\ \citenamefont {Capasso}}]{Devlin:2017-896:SCI}%
  \BibitemOpen
  \bibfield  {author} {\bibinfo {author} {\bibfnamefont {R.~C.}\ \bibnamefont
  {Devlin}}, \bibinfo {author} {\bibfnamefont {A.}~\bibnamefont {Ambrosio}},
  \bibinfo {author} {\bibfnamefont {N.~A.}\ \bibnamefont {Rubin}}, \bibinfo
  {author} {\bibfnamefont {J.~P.~B.}\ \bibnamefont {Mueller}}, \ and\ \bibinfo
  {author} {\bibfnamefont {F.}~\bibnamefont {Capasso}},\ }\bibfield  {title}
  {\enquote {\bibinfo {title} {Arbitrary spin-to{\textendash}orbital angular
  momentum conversion of light},}\ }\href {\doibase 10.1126/science.aao5392}
  {\bibfield  {journal} {\bibinfo  {journal} {Science}\ }\textbf {\bibinfo
  {volume} {358}},\ \bibinfo {pages} {896--901} (\bibinfo {year}
  {2017})}\BibitemShut {NoStop}%
\bibitem [{\citenamefont {Jha}\ \emph {et~al.}(2015)\citenamefont {Jha},
  \citenamefont {Ni}, \citenamefont {Wu}, \citenamefont {Wang},\ and\
  \citenamefont {Zhang}}]{Jha:2015-25501:PRL}%
  \BibitemOpen
  \bibfield  {author} {\bibinfo {author} {\bibfnamefont {P.~K.}\ \bibnamefont
  {Jha}}, \bibinfo {author} {\bibfnamefont {X.~J.}\ \bibnamefont {Ni}},
  \bibinfo {author} {\bibfnamefont {C.~H.}\ \bibnamefont {Wu}}, \bibinfo
  {author} {\bibfnamefont {Y.}~\bibnamefont {Wang}}, \ and\ \bibinfo {author}
  {\bibfnamefont {X.}~\bibnamefont {Zhang}},\ }\bibfield  {title}
  {{\selectlanguage {English}\enquote {\bibinfo {title} {Metasurface-enabled
  remote quantum interference},}\ }}\href {\doibase
  10.1103/PhysRevLett.115.025501} {\bibfield  {journal} {\bibinfo  {journal}
  {Phys. Rev. Lett.}\ }\textbf {\bibinfo {volume} {115}},\ \bibinfo {pages}
  {025501--5} (\bibinfo {year} {2015})}\BibitemShut {NoStop}%
\bibitem [{\citenamefont {Roger}\ \emph {et~al.}(2015)\citenamefont {Roger},
  \citenamefont {Vezzoli}, \citenamefont {Bolduc}, \citenamefont {Valente},
  \citenamefont {Heitz}, \citenamefont {Jeffers}, \citenamefont {Soci},
  \citenamefont {Leach}, \citenamefont {Couteau}, \citenamefont {Zheludev},\
  and\ \citenamefont {Faccio}}]{Roger:2015-7031:NCOM}%
  \BibitemOpen
  \bibfield  {author} {\bibinfo {author} {\bibfnamefont {T.}~\bibnamefont
  {Roger}}, \bibinfo {author} {\bibfnamefont {S.}~\bibnamefont {Vezzoli}},
  \bibinfo {author} {\bibfnamefont {E.}~\bibnamefont {Bolduc}}, \bibinfo
  {author} {\bibfnamefont {J.}~\bibnamefont {Valente}}, \bibinfo {author}
  {\bibfnamefont {J.~J.~F.}\ \bibnamefont {Heitz}}, \bibinfo {author}
  {\bibfnamefont {J.}~\bibnamefont {Jeffers}}, \bibinfo {author} {\bibfnamefont
  {C.}~\bibnamefont {Soci}}, \bibinfo {author} {\bibfnamefont {J.}~\bibnamefont
  {Leach}}, \bibinfo {author} {\bibfnamefont {C.}~\bibnamefont {Couteau}},
  \bibinfo {author} {\bibfnamefont {N.~I.}\ \bibnamefont {Zheludev}}, \ and\
  \bibinfo {author} {\bibfnamefont {D.}~\bibnamefont {Faccio}},\ }\bibfield
  {title} {{\selectlanguage {English}\enquote {\bibinfo {title} {Coherent
  perfect absorption in deeply subwavelength films in the single-photon
  regime},}\ }}\href {\doibase 10.1038/ncomms8031} {\bibfield  {journal}
  {\bibinfo  {journal} {Nat. Commun.}\ }\textbf {\bibinfo {volume} {6}},\
  \bibinfo {pages} {7031--5} (\bibinfo {year} {2015})}\BibitemShut {NoStop}%
\bibitem [{\citenamefont {Lyons}\ \emph {et~al.}(2017)\citenamefont {Lyons},
  \citenamefont {Oren}, \citenamefont {Roger}, \citenamefont {Savinov},
  \citenamefont {Valente}, \citenamefont {Vezzoli}, \citenamefont {Zheludev},
  \citenamefont {Segev},\ and\ \citenamefont
  {Faccio}}]{Lyons:1709.03428:ARXIV}%
  \BibitemOpen
  \bibfield  {author} {\bibinfo {author} {\bibfnamefont {A.}~\bibnamefont
  {Lyons}}, \bibinfo {author} {\bibfnamefont {D.}~\bibnamefont {Oren}},
  \bibinfo {author} {\bibfnamefont {T.}~\bibnamefont {Roger}}, \bibinfo
  {author} {\bibfnamefont {V.}~\bibnamefont {Savinov}}, \bibinfo {author}
  {\bibfnamefont {J.}~\bibnamefont {Valente}}, \bibinfo {author} {\bibfnamefont
  {S.}~\bibnamefont {Vezzoli}}, \bibinfo {author} {\bibfnamefont {N.~I.}\
  \bibnamefont {Zheludev}}, \bibinfo {author} {\bibfnamefont {M.}~\bibnamefont
  {Segev}}, \ and\ \bibinfo {author} {\bibfnamefont {D.}~\bibnamefont
  {Faccio}},\ }\bibfield  {title} {\enquote {\bibinfo {title} {Coherent
  metamaterial absorption of two-photon states with 40\% efficiency},}\ }\href
  {https://arxiv.org/abs/1709.03428} {\bibfield  {journal} {\bibinfo  {journal}
  {arXiv}\ }\textbf {\bibinfo {volume} {\mdseries 1709.03428}} (\bibinfo {year}
  {2017})}\BibitemShut {NoStop}%
\bibitem [{\citenamefont {Stav}\ \emph {et~al.}(2018)\citenamefont {Stav},
  \citenamefont {Faerman}, \citenamefont {Maguid}, \citenamefont {Oren},
  \citenamefont {Kleiner}, \citenamefont {Hasman},\ and\ \citenamefont
  {Segev}}]{Stav:1802.06374:ARXIV}%
  \BibitemOpen
  \bibfield  {author} {\bibinfo {author} {\bibfnamefont {T.}~\bibnamefont
  {Stav}}, \bibinfo {author} {\bibfnamefont {A.}~\bibnamefont {Faerman}},
  \bibinfo {author} {\bibfnamefont {E.}~\bibnamefont {Maguid}}, \bibinfo
  {author} {\bibfnamefont {D.}~\bibnamefont {Oren}}, \bibinfo {author}
  {\bibfnamefont {V.}~\bibnamefont {Kleiner}}, \bibinfo {author} {\bibfnamefont
  {E.}~\bibnamefont {Hasman}}, \ and\ \bibinfo {author} {\bibfnamefont
  {M.}~\bibnamefont {Segev}},\ }\bibfield  {title} {\enquote {\bibinfo {title}
  {Quantum metamaterials: entanglement of spin and orbital angular momentum of
  a single photon},}\ }\href {https://arxiv.org/abs/1802.06374} {\bibfield
  {journal} {\bibinfo  {journal} {arXiv}\ }\textbf {\bibinfo {volume}
  {\mdseries 1802.06374}} (\bibinfo {year} {2018})}\BibitemShut {NoStop}%
\bibitem [{\citenamefont {Silverstone}\ \emph {et~al.}(2014)\citenamefont
  {Silverstone}, \citenamefont {Bonneau}, \citenamefont {Ohira}, \citenamefont
  {Suzuki}, \citenamefont {Yoshida}, \citenamefont {Iizuka}, \citenamefont
  {Ezaki}, \citenamefont {Natarajan}, \citenamefont {Tanner}, \citenamefont
  {Hadfield}, \citenamefont {Zwiller}, \citenamefont {Marshall}, \citenamefont
  {Rarity}, \citenamefont {O'Brien},\ and\ \citenamefont
  {Thompson}}]{Silverstone:2014-104:NPHOT}%
  \BibitemOpen
  \bibfield  {author} {\bibinfo {author} {\bibfnamefont {J.~W.}\ \bibnamefont
  {Silverstone}}, \bibinfo {author} {\bibfnamefont {D.}~\bibnamefont
  {Bonneau}}, \bibinfo {author} {\bibfnamefont {K.}~\bibnamefont {Ohira}},
  \bibinfo {author} {\bibfnamefont {N.}~\bibnamefont {Suzuki}}, \bibinfo
  {author} {\bibfnamefont {H.}~\bibnamefont {Yoshida}}, \bibinfo {author}
  {\bibfnamefont {N.}~\bibnamefont {Iizuka}}, \bibinfo {author} {\bibfnamefont
  {M.}~\bibnamefont {Ezaki}}, \bibinfo {author} {\bibfnamefont {C.~M.}\
  \bibnamefont {Natarajan}}, \bibinfo {author} {\bibfnamefont {M.~G.}\
  \bibnamefont {Tanner}}, \bibinfo {author} {\bibfnamefont {R.~H.}\
  \bibnamefont {Hadfield}}, \bibinfo {author} {\bibfnamefont {V.}~\bibnamefont
  {Zwiller}}, \bibinfo {author} {\bibfnamefont {G.~D.}\ \bibnamefont
  {Marshall}}, \bibinfo {author} {\bibfnamefont {J.~G.}\ \bibnamefont
  {Rarity}}, \bibinfo {author} {\bibfnamefont {J.~L.}\ \bibnamefont {O'Brien}},
  \ and\ \bibinfo {author} {\bibfnamefont {M.~G.}\ \bibnamefont {Thompson}},\
  }\bibfield  {title} {{\selectlanguage {English}\enquote {\bibinfo {title}
  {On-chip quantum interference between silicon photon-pair sources},}\ }}\href
  {\doibase 10.1038/NPHOTON.2013.339} {\bibfield  {journal} {\bibinfo
  {journal} {Nature Photonics}\ }\textbf {\bibinfo {volume} {8}},\ \bibinfo
  {pages} {104--108} (\bibinfo {year} {2014})}\BibitemShut {NoStop}%
\bibitem [{\citenamefont {Bayraktar}\ \emph {et~al.}(2016)\citenamefont
  {Bayraktar}, \citenamefont {Swillo}, \citenamefont {Canalias},\ and\
  \citenamefont {Bjork}}]{Bayraktar:2016-20105:PRA}%
  \BibitemOpen
  \bibfield  {author} {\bibinfo {author} {\bibfnamefont {O.}~\bibnamefont
  {Bayraktar}}, \bibinfo {author} {\bibfnamefont {M.}~\bibnamefont {Swillo}},
  \bibinfo {author} {\bibfnamefont {C.}~\bibnamefont {Canalias}}, \ and\
  \bibinfo {author} {\bibfnamefont {G.}~\bibnamefont {Bjork}},\ }\bibfield
  {title} {{\selectlanguage {English}\enquote {\bibinfo {title}
  {Quantum-polarization state tomography},}\ }}\href {\doibase
  10.1103/PhysRevA.94.020105} {\bibfield  {journal} {\bibinfo  {journal} {Phys.
  Rev. A}\ }\textbf {\bibinfo {volume} {94}},\ \bibinfo {pages} {020105--5}
  (\bibinfo {year} {2016})}\BibitemShut {NoStop}%
\bibitem [{\citenamefont {Fakonas}\ \emph {et~al.}(2015)\citenamefont
  {Fakonas}, \citenamefont {Mitskovets},\ and\ \citenamefont
  {Atwater}}]{Fakonas:2015-23002:NJP}%
  \BibitemOpen
  \bibfield  {author} {\bibinfo {author} {\bibfnamefont {J.~S.}\ \bibnamefont
  {Fakonas}}, \bibinfo {author} {\bibfnamefont {A.}~\bibnamefont {Mitskovets}},
  \ and\ \bibinfo {author} {\bibfnamefont {H.~A.}\ \bibnamefont {Atwater}},\
  }\bibfield  {title} {{\selectlanguage {English}\enquote {\bibinfo {title}
  {Path entanglement of surface plasmons},}\ }}\href {\doibase
  10.1088/1367-2630/17/2/023002} {\bibfield  {journal} {\bibinfo  {journal}
  {New J. Phys.}\ }\textbf {\bibinfo {volume} {17}},\ \bibinfo {pages}
  {023002--7} (\bibinfo {year} {2015})}\BibitemShut {NoStop}%
\bibitem [{\citenamefont {Di~Martino}\ \emph {et~al.}(2014)\citenamefont
  {Di~Martino}, \citenamefont {Sonnefraud}, \citenamefont {Tame}, \citenamefont
  {Kena-Cohen}, \citenamefont {Dieleman}, \citenamefont {Ozdemir},
  \citenamefont {Kim},\ and\ \citenamefont
  {Maier}}]{DiMartino:2014-34004:PRAP}%
  \BibitemOpen
  \bibfield  {author} {\bibinfo {author} {\bibfnamefont {G.}~\bibnamefont
  {Di~Martino}}, \bibinfo {author} {\bibfnamefont {Y.}~\bibnamefont
  {Sonnefraud}}, \bibinfo {author} {\bibfnamefont {M.~S.}\ \bibnamefont
  {Tame}}, \bibinfo {author} {\bibfnamefont {S.}~\bibnamefont {Kena-Cohen}},
  \bibinfo {author} {\bibfnamefont {F.}~\bibnamefont {Dieleman}}, \bibinfo
  {author} {\bibfnamefont {S.~K.}\ \bibnamefont {Ozdemir}}, \bibinfo {author}
  {\bibfnamefont {M.~S.}\ \bibnamefont {Kim}}, \ and\ \bibinfo {author}
  {\bibfnamefont {S.~A.}\ \bibnamefont {Maier}},\ }\bibfield  {title}
  {{\selectlanguage {English}\enquote {\bibinfo {title} {Observation of quantum
  interference in the plasmonic {H}ong-{O}u-{M}andel effect},}\ }}\href
  {\doibase 10.1103/PhysRevApplied.1.034004} {\bibfield  {journal} {\bibinfo
  {journal} {Phys. Rev. Appl.}\ }\textbf {\bibinfo {volume} {1}},\ \bibinfo
  {pages} {034004--6} (\bibinfo {year} {2014})}\BibitemShut {NoStop}%
\bibitem [{\citenamefont {Vest}\ \emph {et~al.}(2017)\citenamefont {Vest},
  \citenamefont {Dheur}, \citenamefont {Devaux}, \citenamefont {Baron},
  \citenamefont {Rousseau}, \citenamefont {Hugonin}, \citenamefont {Greffet},
  \citenamefont {Messin},\ and\ \citenamefont {Marquier}}]{Vest:2017-1373:SCI}%
  \BibitemOpen
  \bibfield  {author} {\bibinfo {author} {\bibfnamefont {B.}~\bibnamefont
  {Vest}}, \bibinfo {author} {\bibfnamefont {M.~C.}\ \bibnamefont {Dheur}},
  \bibinfo {author} {\bibfnamefont {E.}~\bibnamefont {Devaux}}, \bibinfo
  {author} {\bibfnamefont {A.}~\bibnamefont {Baron}}, \bibinfo {author}
  {\bibfnamefont {E.}~\bibnamefont {Rousseau}}, \bibinfo {author}
  {\bibfnamefont {J.~P.}\ \bibnamefont {Hugonin}}, \bibinfo {author}
  {\bibfnamefont {J.~J.}\ \bibnamefont {Greffet}}, \bibinfo {author}
  {\bibfnamefont {G.}~\bibnamefont {Messin}}, \ and\ \bibinfo {author}
  {\bibfnamefont {F.}~\bibnamefont {Marquier}},\ }\bibfield  {title}
  {{\selectlanguage {English}\enquote {\bibinfo {title} {Anti-coalescence of
  bosons on a lossy beam splitter},}\ }}\href {\doibase
  10.1126/science.aam9353} {\bibfield  {journal} {\bibinfo  {journal}
  {Science}\ }\textbf {\bibinfo {volume} {356}},\ \bibinfo {pages} {1373--1376}
  (\bibinfo {year} {2017})}\BibitemShut {NoStop}%
\bibitem [{\citenamefont {Foreman}\ \emph {et~al.}(2015)\citenamefont
  {Foreman}, \citenamefont {Favaro},\ and\ \citenamefont
  {Aiello}}]{Foreman:2015-263901:PRL}%
  \BibitemOpen
  \bibfield  {author} {\bibinfo {author} {\bibfnamefont {M.~R.}\ \bibnamefont
  {Foreman}}, \bibinfo {author} {\bibfnamefont {A.}~\bibnamefont {Favaro}}, \
  and\ \bibinfo {author} {\bibfnamefont {A.}~\bibnamefont {Aiello}},\
  }\bibfield  {title} {{\selectlanguage {English}\enquote {\bibinfo {title}
  {Optimal frames for polarization state reconstruction},}\ }}\href {\doibase
  10.1103/PhysRevLett.115.263901} {\bibfield  {journal} {\bibinfo  {journal}
  {Phys. Rev. Lett.}\ }\textbf {\bibinfo {volume} {115}},\ \bibinfo {pages}
  {263901--6} (\bibinfo {year} {2015})}\BibitemShut {NoStop}%
\bibitem [{\citenamefont {Edgar}\ \emph {et~al.}(2012)\citenamefont {Edgar},
  \citenamefont {Tasca}, \citenamefont {Izdebski}, \citenamefont {Warburton},
  \citenamefont {Leach}, \citenamefont {Agnew}, \citenamefont {Buller},
  \citenamefont {Boyd},\ and\ \citenamefont {Padgett}}]{Edgar:2012-984:NCOM}%
  \BibitemOpen
  \bibfield  {author} {\bibinfo {author} {\bibfnamefont {M.~P.}\ \bibnamefont
  {Edgar}}, \bibinfo {author} {\bibfnamefont {D.~S.}\ \bibnamefont {Tasca}},
  \bibinfo {author} {\bibfnamefont {F.}~\bibnamefont {Izdebski}}, \bibinfo
  {author} {\bibfnamefont {R.~E.}\ \bibnamefont {Warburton}}, \bibinfo {author}
  {\bibfnamefont {J.}~\bibnamefont {Leach}}, \bibinfo {author} {\bibfnamefont
  {M.}~\bibnamefont {Agnew}}, \bibinfo {author} {\bibfnamefont {G.~S.}\
  \bibnamefont {Buller}}, \bibinfo {author} {\bibfnamefont {R.~W.}\
  \bibnamefont {Boyd}}, \ and\ \bibinfo {author} {\bibfnamefont {M.~J.}\
  \bibnamefont {Padgett}},\ }\bibfield  {title} {{\selectlanguage
  {English}\enquote {\bibinfo {title} {Imaging high-dimensional spatial
  entanglement with a camera},}\ }}\href {\doibase 10.1038/ncomms1988}
  {\bibfield  {journal} {\bibinfo  {journal} {Nat. Commun.}\ }\textbf {\bibinfo
  {volume} {3}},\ \bibinfo {pages} {984--6} (\bibinfo {year}
  {2012})}\BibitemShut {NoStop}%
\bibitem [{\citenamefont {Reichert}\ \emph {et~al.}(2018)\citenamefont
  {Reichert}, \citenamefont {Defienne},\ and\ \citenamefont
  {Fleischer}}]{Reichert:2018-7925:SRP}%
  \BibitemOpen
  \bibfield  {author} {\bibinfo {author} {\bibfnamefont {M.}~\bibnamefont
  {Reichert}}, \bibinfo {author} {\bibfnamefont {H.}~\bibnamefont {Defienne}},
  \ and\ \bibinfo {author} {\bibfnamefont {J.~W.}\ \bibnamefont {Fleischer}},\
  }\bibfield  {title} {{\selectlanguage {English}\enquote {\bibinfo {title}
  {Massively parallel coincidence counting of high-dimensional entangled
  states},}\ }}\href {\doibase 10.1038/s41598-018-26144-7} {\bibfield
  {journal} {\bibinfo  {journal} {Sci. Rep.}\ }\textbf {\bibinfo {volume}
  {8}},\ \bibinfo {pages} {7925--7} (\bibinfo {year} {2018})}\BibitemShut
  {NoStop}%
\bibitem [{\citenamefont {James}\ \emph {et~al.}(2001)\citenamefont {James},
  \citenamefont {Kwiat}, \citenamefont {Munro},\ and\ \citenamefont
  {White}}]{James:2001-52312:PRA}%
  \BibitemOpen
  \bibfield  {author} {\bibinfo {author} {\bibfnamefont {D.~F.~V.}\
  \bibnamefont {James}}, \bibinfo {author} {\bibfnamefont {P.~G.}\ \bibnamefont
  {Kwiat}}, \bibinfo {author} {\bibfnamefont {W.~J.}\ \bibnamefont {Munro}}, \
  and\ \bibinfo {author} {\bibfnamefont {A.~G.}\ \bibnamefont {White}},\
  }\bibfield  {title} {{\selectlanguage {English}\enquote {\bibinfo {title}
  {Measurement of qubits},}\ }}\href {\doibase 10.1103/PhysRevA.64.052312}
  {\bibfield  {journal} {\bibinfo  {journal} {Phys. Rev. A}\ }\textbf {\bibinfo
  {volume} {64}},\ \bibinfo {pages} {052312--15} (\bibinfo {year}
  {2001})}\BibitemShut {NoStop}%
\bibitem [{\citenamefont {Titchener}\ \emph {et~al.}(2016)\citenamefont
  {Titchener}, \citenamefont {Solntsev},\ and\ \citenamefont
  {Sukhorukov}}]{Titchener:2016-4079:OL}%
  \BibitemOpen
  \bibfield  {author} {\bibinfo {author} {\bibfnamefont {J.~G.}\ \bibnamefont
  {Titchener}}, \bibinfo {author} {\bibfnamefont {A.~S.}\ \bibnamefont
  {Solntsev}}, \ and\ \bibinfo {author} {\bibfnamefont {A.~A.}\ \bibnamefont
  {Sukhorukov}},\ }\bibfield  {title} {{\selectlanguage {English}\enquote
  {\bibinfo {title} {Two-photon tomography using on-chip quantum walks},}\
  }}\href {\doibase 10.1364/OL.41.004079} {\bibfield  {journal} {\bibinfo
  {journal} {Opt. Lett.}\ }\textbf {\bibinfo {volume} {41}},\ \bibinfo {pages}
  {4079--4082} (\bibinfo {year} {2016})}\BibitemShut {NoStop}%
\bibitem [{\citenamefont {Oren}\ \emph {et~al.}(2017)\citenamefont {Oren},
  \citenamefont {Mutzafi}, \citenamefont {Eldar},\ and\ \citenamefont
  {Segev}}]{Oren:2017-993:OPT}%
  \BibitemOpen
  \bibfield  {author} {\bibinfo {author} {\bibfnamefont {D.}~\bibnamefont
  {Oren}}, \bibinfo {author} {\bibfnamefont {M.}~\bibnamefont {Mutzafi}},
  \bibinfo {author} {\bibfnamefont {Y.~C.}\ \bibnamefont {Eldar}}, \ and\
  \bibinfo {author} {\bibfnamefont {M.}~\bibnamefont {Segev}},\ }\bibfield
  {title} {{\selectlanguage {English}\enquote {\bibinfo {title} {Quantum state
  tomography with a single measurement setup},}\ }}\href {\doibase
  10.1364/OPTICA.4.000993} {\bibfield  {journal} {\bibinfo  {journal} {Optica}\
  }\textbf {\bibinfo {volume} {4}},\ \bibinfo {pages} {993--999} (\bibinfo
  {year} {2017})}\BibitemShut {NoStop}%
\bibitem [{\citenamefont {Titchener}\ \emph {et~al.}(2018)\citenamefont
  {Titchener}, \citenamefont {Gr{\"{a}}fe}, \citenamefont {Heilmann},
  \citenamefont {Solntsev}, \citenamefont {Szameit},\ and\ \citenamefont
  {Sukhorukov}}]{Titchener:2018-19:NPJQI}%
  \BibitemOpen
  \bibfield  {author} {\bibinfo {author} {\bibfnamefont {J.~G.}\ \bibnamefont
  {Titchener}}, \bibinfo {author} {\bibfnamefont {M.}~\bibnamefont
  {Gr{\"{a}}fe}}, \bibinfo {author} {\bibfnamefont {R.}~\bibnamefont
  {Heilmann}}, \bibinfo {author} {\bibfnamefont {A.~S.}\ \bibnamefont
  {Solntsev}}, \bibinfo {author} {\bibfnamefont {A.}~\bibnamefont {Szameit}}, \
  and\ \bibinfo {author} {\bibfnamefont {A.~A.}\ \bibnamefont {Sukhorukov}},\
  }\bibfield  {title} {\enquote {\bibinfo {title} {Scalable on-chip quantum
  state tomography},}\ }\href {\doibase 10.1038/s41534-018-0063-5} {\bibfield
  {journal} {\bibinfo  {journal} {npj Quant. Inform.}\ }\textbf {\bibinfo
  {volume} {4}},\ \bibinfo {pages} {19} (\bibinfo {year} {2018})}\BibitemShut
  {NoStop}%
\bibitem [{\citenamefont {Shadbolt}\ \emph {et~al.}(2012)\citenamefont
  {Shadbolt}, \citenamefont {Verde}, \citenamefont {Peruzzo}, \citenamefont
  {Politi}, \citenamefont {Laing}, \citenamefont {Lobino}, \citenamefont
  {Matthews}, \citenamefont {Thompson},\ and\ \citenamefont
  {O'Brien}}]{Shadbolt:2012-45:NPHOT}%
  \BibitemOpen
  \bibfield  {author} {\bibinfo {author} {\bibfnamefont {P.~J.}\ \bibnamefont
  {Shadbolt}}, \bibinfo {author} {\bibfnamefont {M.~R.}\ \bibnamefont {Verde}},
  \bibinfo {author} {\bibfnamefont {A.}~\bibnamefont {Peruzzo}}, \bibinfo
  {author} {\bibfnamefont {A.}~\bibnamefont {Politi}}, \bibinfo {author}
  {\bibfnamefont {A.}~\bibnamefont {Laing}}, \bibinfo {author} {\bibfnamefont
  {M.}~\bibnamefont {Lobino}}, \bibinfo {author} {\bibfnamefont {J.~C.~F.}\
  \bibnamefont {Matthews}}, \bibinfo {author} {\bibfnamefont {M.~G.}\
  \bibnamefont {Thompson}}, \ and\ \bibinfo {author} {\bibfnamefont {J.~L.}\
  \bibnamefont {O'Brien}},\ }\bibfield  {title} {{\selectlanguage
  {English}\enquote {\bibinfo {title} {Generating, manipulating and measuring
  entanglement and mixture with a reconfigurable photonic circuit},}\ }}\href
  {\doibase 10.1038/NPHOTON.2011.283} {\bibfield  {journal} {\bibinfo
  {journal} {Nature Photonics}\ }\textbf {\bibinfo {volume} {6}},\ \bibinfo
  {pages} {45--49} (\bibinfo {year} {2012})}\BibitemShut {NoStop}%
\bibitem [{\citenamefont {Hong}\ \emph {et~al.}(1987)\citenamefont {Hong},
  \citenamefont {Ou},\ and\ \citenamefont {Mandel}}]{Hong:1987-2044:PRL}%
  \BibitemOpen
  \bibfield  {author} {\bibinfo {author} {\bibfnamefont {C.~K.}\ \bibnamefont
  {Hong}}, \bibinfo {author} {\bibfnamefont {Z.~Y.}\ \bibnamefont {Ou}}, \ and\
  \bibinfo {author} {\bibfnamefont {L.}~\bibnamefont {Mandel}},\ }\bibfield
  {title} {{\selectlanguage {English}\enquote {\bibinfo {title} {Measurement of
  subpicosecond time intervals between 2 photons by interference},}\ }}\href
  {\doibase 10.1103/PhysRevLett.59.2044} {\bibfield  {journal} {\bibinfo
  {journal} {Phys. Rev. Lett.}\ }\textbf {\bibinfo {volume} {59}},\ \bibinfo
  {pages} {2044--2046} (\bibinfo {year} {1987})}\BibitemShut {NoStop}%
\bibitem [{\citenamefont {Maguid}\ \emph {et~al.}(2017)\citenamefont {Maguid},
  \citenamefont {Yulevich}, \citenamefont {Yannai}, \citenamefont {Kleiner},
  \citenamefont {Brongersma},\ and\ \citenamefont
  {Hasman}}]{Maguid:2017-e17027:LSA}%
  \BibitemOpen
  \bibfield  {author} {\bibinfo {author} {\bibfnamefont {E.}~\bibnamefont
  {Maguid}}, \bibinfo {author} {\bibfnamefont {I.}~\bibnamefont {Yulevich}},
  \bibinfo {author} {\bibfnamefont {M.}~\bibnamefont {Yannai}}, \bibinfo
  {author} {\bibfnamefont {V.}~\bibnamefont {Kleiner}}, \bibinfo {author}
  {\bibfnamefont {M.~L.}\ \bibnamefont {Brongersma}}, \ and\ \bibinfo {author}
  {\bibfnamefont {E.}~\bibnamefont {Hasman}},\ }\bibfield  {title}
  {{\selectlanguage {English}\enquote {\bibinfo {title} {Multifunctional
  interleaved geometric-phase dielectric metasurfaces},}\ }}\href {\doibase
  10.1038/lsa.2017.27} {\bibfield  {journal} {\bibinfo  {journal} {Light-Sci.
  Appl.}\ }\textbf {\bibinfo {volume} {6}},\ \bibinfo {pages} {e17027--7}
  (\bibinfo {year} {2017})}\BibitemShut {NoStop}%
\bibitem [{\citenamefont {Myers}\ \emph {et~al.}(1995)\citenamefont {Myers},
  \citenamefont {Eckardt}, \citenamefont {Fejer}, \citenamefont {Byer},
  \citenamefont {Bosenberg},\ and\ \citenamefont
  {Pierce}}]{Myers:1995-2102:JOSB}%
  \BibitemOpen
  \bibfield  {author} {\bibinfo {author} {\bibfnamefont {L.~E.}\ \bibnamefont
  {Myers}}, \bibinfo {author} {\bibfnamefont {R.~C.}\ \bibnamefont {Eckardt}},
  \bibinfo {author} {\bibfnamefont {M.~M.}\ \bibnamefont {Fejer}}, \bibinfo
  {author} {\bibfnamefont {R.~L.}\ \bibnamefont {Byer}}, \bibinfo {author}
  {\bibfnamefont {W.~R.}\ \bibnamefont {Bosenberg}}, \ and\ \bibinfo {author}
  {\bibfnamefont {J.~W.}\ \bibnamefont {Pierce}},\ }\bibfield  {title}
  {{\selectlanguage {English}\enquote {\bibinfo {title} {Quasi-phase-matched
  optical parametric oscillators in bulk periodically poled {{LiNbO}$_3$}},}\
  }}\href {http://www.opticsinfobase.org/abstract.cfm?URI=josab-12-11-2102}
  {\bibfield  {journal} {\bibinfo  {journal} {J. Opt. Soc. Am. B}\ }\textbf
  {\bibinfo {volume} {12}},\ \bibinfo {pages} {2102--2116} (\bibinfo {year}
  {1995})}\BibitemShut {NoStop}%
\bibitem [{\citenamefont {Chung}\ \emph {et~al.}(2015)\citenamefont {Chung},
  \citenamefont {Huang}, \citenamefont {Yang}, \citenamefont {Chang},
  \citenamefont {Wu}, \citenamefont {Setzpfandt}, \citenamefont {Pertsch},
  \citenamefont {Neshev},\ and\ \citenamefont {Chen}}]{Chung:2015-30641:OE}%
  \BibitemOpen
  \bibfield  {author} {\bibinfo {author} {\bibfnamefont {H.~P.}\ \bibnamefont
  {Chung}}, \bibinfo {author} {\bibfnamefont {K.~H.}\ \bibnamefont {Huang}},
  \bibinfo {author} {\bibfnamefont {S.~L.}\ \bibnamefont {Yang}}, \bibinfo
  {author} {\bibfnamefont {W.~K.}\ \bibnamefont {Chang}}, \bibinfo {author}
  {\bibfnamefont {C.~W.}\ \bibnamefont {Wu}}, \bibinfo {author} {\bibfnamefont
  {F.}~\bibnamefont {Setzpfandt}}, \bibinfo {author} {\bibfnamefont
  {T.}~\bibnamefont {Pertsch}}, \bibinfo {author} {\bibfnamefont {D.~N.}\
  \bibnamefont {Neshev}}, \ and\ \bibinfo {author} {\bibfnamefont {Y.~H.}\
  \bibnamefont {Chen}},\ }\bibfield  {title} {{\selectlanguage
  {English}\enquote {\bibinfo {title} {Adiabatic light transfer in titanium
  diffused lithium niobate waveguides},}\ }}\href {\doibase
  10.1364/OE.23.030641} {\bibfield  {journal} {\bibinfo  {journal} {Opt.
  Express}\ }\textbf {\bibinfo {volume} {23}},\ \bibinfo {pages} {30641--30650}
  (\bibinfo {year} {2015})}\BibitemShut {NoStop}%
\bibitem [{\citenamefont {Lenzini}\ \emph {et~al.}(2018)\citenamefont
  {Lenzini}, \citenamefont {Poddubny}, \citenamefont {Titchener}, \citenamefont
  {Fisher}, \citenamefont {Boes}, \citenamefont {Kasture}, \citenamefont
  {Haylock}, \citenamefont {Villa}, \citenamefont {Mitchell}, \citenamefont
  {Solntsev}, \citenamefont {Sukhorukov},\ and\ \citenamefont
  {Lobino}}]{Lenzini:2018-17143:LSA}%
  \BibitemOpen
  \bibfield  {author} {\bibinfo {author} {\bibfnamefont {F.}~\bibnamefont
  {Lenzini}}, \bibinfo {author} {\bibfnamefont {A.~N.}\ \bibnamefont
  {Poddubny}}, \bibinfo {author} {\bibfnamefont {J.}~\bibnamefont {Titchener}},
  \bibinfo {author} {\bibfnamefont {P.}~\bibnamefont {Fisher}}, \bibinfo
  {author} {\bibfnamefont {A.}~\bibnamefont {Boes}}, \bibinfo {author}
  {\bibfnamefont {S.}~\bibnamefont {Kasture}}, \bibinfo {author} {\bibfnamefont
  {B.}~\bibnamefont {Haylock}}, \bibinfo {author} {\bibfnamefont
  {M.}~\bibnamefont {Villa}}, \bibinfo {author} {\bibfnamefont
  {A.}~\bibnamefont {Mitchell}}, \bibinfo {author} {\bibfnamefont {A.~S.}\
  \bibnamefont {Solntsev}}, \bibinfo {author} {\bibfnamefont {A.~A.}\
  \bibnamefont {Sukhorukov}}, \ and\ \bibinfo {author} {\bibfnamefont
  {M.}~\bibnamefont {Lobino}},\ }\bibfield  {title} {{\selectlanguage
  {English}\enquote {\bibinfo {title} {Direct characterization of a nonlinear
  photonic circuit's wave function with laser light},}\ }}\href {\doibase
  10.1038/lsa.2017.143} {\bibfield  {journal} {\bibinfo  {journal} {Light-Sci.
  Appl.}\ }\textbf {\bibinfo {volume} {7}},\ \bibinfo {pages} {17143--5}
  (\bibinfo {year} {2018})}\BibitemShut {NoStop}%
\bibitem [{\citenamefont {Wootters}(1998)}]{Wootters:1998-2245:PRL}%
  \BibitemOpen
  \bibfield  {author} {\bibinfo {author} {\bibfnamefont {W.~K.}\ \bibnamefont
  {Wootters}},\ }\bibfield  {title} {{\selectlanguage {English}\enquote
  {\bibinfo {title} {Entanglement of formation of an arbitrary state of two
  qubits},}\ }}\href {\doibase 10.1103/PhysRevLett.80.2245} {\bibfield
  {journal} {\bibinfo  {journal} {Phys. Rev. Lett.}\ }\textbf {\bibinfo
  {volume} {80}},\ \bibinfo {pages} {2245--2248} (\bibinfo {year}
  {1998})}\BibitemShut {NoStop}%
\end{thebibliography}
\end{document}